\documentclass[useAMS,usenatbib,usegraphicx]{mn2e}

\usepackage{amsmath}
\usepackage{float}
\usepackage[font=footnotesize,caption=false]{subfig}
\usepackage{booktabs}

\newenvironment{squishitemize}
{\begin{list}{\textbullet}{%
    \setlength{\itemsep}{0pt}%
    \setlength{\parsep}{0pt}%
    \setlength{\topsep}{0pt}%
    \setlength{\parskip}{0pt} %
    \setlength{\labelwidth}{.5in}%
    \setlength{\labelsep}{0.05in} %
    \setlength{\leftmargin}{.25in} %
    }}
  {\end{list}}

\newcommand{\oracdr}{\small{ORAC-DR }\normalsize{}}

\newcommand{\nar}{New A Rev.}
\newcommand{\mnras}{MNRAS}
\newcommand{\nat}{Nat}
\newcommand{\aj}{AJ}
\newcommand{\apj}{ApJ}
\newcommand{\apjl}{ApJL}
\newcommand{\apjs}{ApJS}

\newcommand{\aap}{A\&A}
\newcommand{\aaps}{A{\&}AS}
\newcommand{\pasp}{PASP}
\newcommand{\araa}{ARA\&A}

\def\Msun{\hbox{M$_{\odot}$}}
\def\sfrunits{\Msun\ yr$^{-1}$}
\def\halpha{H$\alpha$  }
\def\halphans{H$\alpha$}
\def\hbeta{H$\beta$  }
\def\smean{$\sigma_{\mathrm{mean}}$}
\def\vshear{$v_{\mathrm{shear}}$  }
\def\NII{[N\hspace{.03cm}II]}
\def\SII{[S\hspace{.03cm}II]}
\def\OII{[O\hspace{.03cm}II]}
\def\OIII{[O\hspace{.03cm}III]}

\title[WIGGLEZ LUMINOUS STAR-FORMING GALAXIES]{The WiggleZ Dark Energy Survey: High Resolution Kinematics of Luminous Star-Forming Galaxies}
\author[Wisnioski et al.]{Emily Wisnioski$^{1}$\thanks{E-mail: ewisnios@astro.swin.edu.au},
Karl Glazebrook$^{1}$, 
Chris Blake$^{1}$, 
Ted Wyder$^{2}$,
Chris Martin$^{2}$, \newauthor
Gregory B.\ Poole$^{1}$, 
Rob Sharp$^{3}$, 
Warrick Couch$^{1}$, 
Glenn G. Kacprzak$^{1}$,
Sarah \newauthor Brough$^{4}$, 
Matthew Colless$^{4}$, 
Carlos Contreras$^{1}$, 
Scott Croom$^{5}$, 
Darren Croton$^{1}$,\newauthor
Tamara Davis$^{6}$, 
Michael J.\ Drinkwater$^{6}$, 
Karl Forster$^{2}$,
David G. Gilbank$^{7}$, \newauthor
Michael Gladders$^{8}$, 
Ben Jelliffe$^{5}$, 
Russell J.\ Jurek$^{9}$, 
I-hui Li$^{1}$, 
Barry Madore$^{10}$, \newauthor
Kevin Pimbblet$^{11}$, 
Michael Pracy$^{1,3}$ 
David Woods$^{12}$ and 
H.K.C. Yee$^{13}$\\
$^{1}$Centre for Astrophysics and Supercomputing, Swinburne University of Technology, P.O. Box 218, Hawthorn, VIC 3122, Australia\\
$^{2}$California Institute of Technology, MC 405-47, 1200 East California Boulevard, Pasadena, CA 91125, United States\\
$^{3}$Research School of Astronomy \& Astrophysics, Australian National University, Weston Creek, ACT 2600, Australia\\
$^{4}$Australian Astronomical Observatory, P.O. Box 296, Epping, NSW 2121, Australia\\
$^{5}$School of Physics, University of Sydney, NSW 2006, Australia\\
$^{6}$Department of Physics, University of Queensland, Brisbane, QLD 4072, Australia\\
$^{7}$Department of Physics and Astronomy, University Of Waterloo, Waterloo, Ontario N2L 3G1, Canada\\
$^{8}$Department of Astronomy and Astrophysics, The University of Chicago, 5640 S. Ellis Ave, Chicago, IL 60637, United States\\
$^{9}$CSIRO Astronomy \& Space Sciences, Australia Telescope National Facility, Epping, NSW, 1710, Australia\\
$^{10}$Observatories of the Carnegie Institute of Washington, 813 Santa Barbara St., Pasadena, CA 91101, United States\\
$^{11}$School of Physics, Monash University, Clayton, VIC 3800, Australia\\
$^{12}$Department of Physics \& Astronomy, University of British Columbia, 6224 Agricultural Road, Vancouver, B.C., V6T 1Z1, Canada\\
$^{13}$Department of Astronomy \& Astrophysics, University of Toronto, 50 St. George St., Toronto, ON, M5S 3H4, Canada}
\begin{document}

\date{{Submitted} 2011 April 30. { Accepted}  2011 July 13.}

\pagerange{\pageref{firstpage}--\pageref{lastpage}} \pubyear{2011}

\maketitle

\label{firstpage}

\begin{abstract}
We report evidence of ordered orbital motion in luminous star-forming galaxies at $z\sim1.3$. We present integral field spectroscopy (IFS) observations, performed with the OH Suppressing InfraRed Imaging Spectrograph (OSIRIS) system, assisted by laser guide star adaptive optics on the Keck telescope, of 13 star-forming galaxies selected from the WiggleZ Dark Energy Survey. Selected via ultraviolet and \OII~emission, the large volume of the WiggleZ survey allows the selection of sources which have comparable intrinsic luminosity and stellar mass to IFS samples at $z>2$.  Multiple 1--2 kpc size sub-components of emission, or `clumps', are detected within the \halpha spatial emission which extends over 6--10 kpc in 4 galaxies, resolved compact emission ($r<3$ kpc) is detected in 5 galaxies, and extended regions of \halpha emission are observed in the remaining 4 galaxies. We discuss these data in the context of different snapshots in a merger sequence and/or the evolutionary stages of coalescence of star-forming regions in an unstable disk. We find evidence of ordered orbital motion in galaxies as expected from disk models and the highest values of velocity dispersion ($\sigma>100$ km s$^{-1}$) in the most compact sources. This unique data set reveals that the most luminous star-forming galaxies at $z>1$ are gaseous unstable disks indicating that a different mode of star formation could be feeding gas to galaxies at $z>1$, and lending support to theories of cold dense gas flows from the intergalactic medium.\\
\end{abstract}

\clearpage

\section{Introduction}
An open question remains regarding the accretion of gas and build up of stars in massive galaxies at the peak epoch of cosmic star formation. The traditional theories of isolated disk-formation resulting from gas accretion from the halo \citep{1962ApJ...136..748E} have been challenged by recent observations and theories at high redshift. In new models of disk formation at high redshift, cooling flows supply gas directly to the centres of galaxies \citep{2005MNRAS.363....2K,Dekel:2009ys}. Direct observations of the gas motions in galaxies are crucial to test these new models and ultimately to disentangle how galaxies form at high redshift.

Morphological studies made possible by Hubble Advanced Camera for Surveys (ACS), Wide Field and Planetary Camera (WFPC) and other deep imagers have provided the first clues to the processes governing high-redshift galaxy formation. High-redshift galaxies exhibit different morphologies to local galaxies, with higher fractions of both compact and irregular galaxies \citep{Glazebrook:1995vn, 1996MNRAS.279L..47A}. This deviation from the Hubble Sequence indicates that different mechanisms may be driving star formation and galaxy evolution at earlier cosmic times. Initial interpretations of the complex structures seen in the Hubble Ultra Deep Field (HUDF) favor higher rates of major mergers, a natural consequence of hierarchical merging models in a $\Lambda$CDM Universe (e.g., \citealt{2003AJ....126.1183C, 2006ApJ...636..592L,  2008ApJ...677...37O,  Engel:2010fk}). 
Yet, despite large dense regions in these galaxies, observations of their stellar populations reveal galactic properties reminiscent of local disk galaxies \citep{2005ApJ...627..632E}. In this case the large dense regions or `clumps' of star formation may be related to HII regions found in spiral galaxies but of much larger size \citep{Jones:2010uf,2010Natur.464..733S}. Models have been developed to explain the clumpy structures in disks and their evolution into galactic structures seen at $z=0$ (e.g., \citealt{1999ApJ...514...77N, 2004A&A...413..547I, 2007ApJ...670..237B, Elmegreen:2008fk}). In these models, turbulent gas collapses from Jeans instabilities resulting in massive star-clusters that, on a timescale of 1 Gyr, coalesce to the centre of the galaxy to form a pseudo-bulge. 

Morphological studies alone do not provide enough information to distinguish between the two kinematically distinct formation theories of merging and isolated disk formation at these redshifts. Given the complex structure of the galaxies it is difficult to derive the major axis for slit positioning making spectroscopic observations less effective in discerning rotation signatures (e.g. \citealt{2004ApJ...612..122E,2006ApJ...646..107E}). Spatially-resolved spectroscopy is therefore essential for disentangling kinematics and star-formation histories of galaxies at high redshift. 

The past five years have seen a great number of integral field spectroscopy (IFS) studies of the nebular emission line, \halphans, in $z>2$ star-forming galaxies (e.g., \citealt{2006Natur.442..786G, 2008ApJ...687...59G,2009ApJ...706.1364F, 2009ApJ...697.2057L, 2007ApJ...658...78W,2009ApJ...699..421W, 2009A&A...504..789E, 2008A&A...479...67N}). These redshifts are favored due to the large number of Lyman break galaxies (LBGs) and sub-millimeter galaxies (SMGs) being discovered by methods tailored for high redshift as well as the accessibility of \halpha in the near-infrared (NIR). 

\begin{table*}
\begin{minipage}{\textwidth}
\caption{OSIRIS observations of WiggleZ galaxies}
\begin{tabular*}{\textwidth}{@{\extracolsep{\fill}}llrrcccccc}
\hline
{WiggleZ ID} & {ID} & {R.A.	} & {Dec. 	} & {z$^{a}$	} & {Obs.} & {$t_{\mathrm{exp.}}$    } & {Exp.    } & {Filter	} & {Scale	} \\
{} & {} & {(J2000.0)	} & {(J2000.0)	} & {} & {Date} & {(s)} & {Seq. } & {} & {(mas)} \\
\vspace{0.25mm} \\
\hline
\multicolumn{10}{c}{Detections}\\
\hline
\vspace{0.25mm} \\
R03J032450240$-$13550943	& WK0912\_13R &  03:24:50.240  & $-$13:55:09.429  &  1.288	&  2009/12	& 3600 & 4 $\times$ 900 &  Hn1	& 50  \\				
S15J145355248$-$00320351	& WK1002\_61S &  14:53:55.247  & $-$00:32:03.501  &  1.305  &  2010/02  & 5400 & 6 $\times$ 900 &  Hn1  & 50 \\
R01J005822757$-$03034040	& WK0909\_02R &  00:58:22.757  & $-$03:03:40.400  &  1.362	&  2009/09	& 5400 & 9 $\times$ 600 &  Hn2	& 50  \\
R03J032206214$-$15443471	& WK0909\_06R &  03:22:06.214  & $-$15:44:34.711  &  1.461	&  2009/09	& 8100 & 9 $\times$ 900 &  Hn3	& 50  \\

S15J144102444+05480354	& WK0905\_22S &  14:41:02.443  &  05:48:03.540  &  1.282	&  2009/05	& 3600 & 4 $\times$ 900 &  Hn1	& 50  \\
R00J232217805$-$05473356	& WK0912\_01R &  23:22:17.805  & $-$05:47:33.560  &  1.297	&  2009/12	& 3600 & 4 $\times$ 900 &  Hn1	& 50  \\				
S09J091517481+00033557	& WK1002\_41S &  09:15:17.483  &  00:03:35.570  &  1.334  &  2010/02  & 7200 & 8 $\times$ 900 &  Hn1  & 50  \\
S00J233338383$-$01040629	& WK0809\_02S &  23:33:38.386  & $-$01:04:06.290  &  1.452	&  2008/09	& 5400 & 6 $\times$ 900 &  Hn3	& 50  \\

S11J101757445$-$00244002	& WK1002\_14S &  10:17:57.444  & $-$00:24:40.020  &  1.305  &  2010/02  & 5400 & 6 $\times$ 900 &  Hn1  & 50  \\
S15J142538641+00483135	& WK1002\_18S &  14:25:38.639  &  00:48:31.350  &  1.305	&  2010/02  & 2700 & 3 $\times$ 900 &  Hn1  & 50  \\
S09J090933680+01074587	& WK1002\_46S &  09:09:33.680  &  01:07:45.870  &  1.308	&  2010/02  & 3000 & 5 $\times$ 600 &  Hn1  & 50  \\
S11J110405504+01185565	& WK1002\_52S &  11:04:05.504  &  01:18:55.650  &  1.320 	&  2010/02  & 7200 & 8 $\times$ 900 &  Hn1  & 50  \\
S09J090312056$-$00273273	& WK0912\_16S &  09:03:12.056  & $-$00:27:32.730  &  1.327	&  2009/12	& 8100 & 9 $\times$ 900 &  Hn1	& 50  \\

R00J233618953$-$10580749$^{b}$	& WK0809\_05R &	 23:36:18.955 & $-$10:58:07.321     &  1.402	&  2008/09	& 2700 & 3 $\times$ 900 &  Hn2 & 100 \\
R00J232248993$-$12151943$^{b}$	& WK0809\_04R &  23:22:48.992  & $-$12:15:19.439  &  1.455	&  2008/09 	& 2700 & 3 $\times$ 900 &  Hn3 & 50  \\
\hline
\vspace{0.25mm} \\
\multicolumn{10}{c}{Mis-Identifications}\\
\hline
\vspace{0.25mm} \\
R22J214700239+01333423	& WK0809\_01R &  21:47:00.239  & 01:33:34.230 &  1.466	&  09/2008	& 3600 & 4 $\times$ 900 &  Hn3	& 50  \\
S00J234420801$-$09322580	& WK0809\_03S &  23:44:20.801  &	$-$09:32:25.800 &  1.488	&  09/2008	& 2700 & 3 $\times$ 900 &  Hn3	& 50  \\
\hline
\label{osirisobs.table}
\end{tabular*} \\
$^{a}$WiggleZ Dark Energy Survey spectroscopic redshift.\\
$^{b}$\halpha detected but observations unusable due to poor AO correction as a result of stray light on the low bandwidth wavefront sensor.
\end{minipage}
\end{table*}

A main focus of these studies is the classification of galaxies kinematically to define clear observational distinctions between disks and interacting systems. Mergers observed with integral field units (IFUs; \citealt{2008A&A...479...67N,2009ApJ...699..421W}) often show clear kinematic steps across one resolution element, indicating separation between two kinematically separate systems, 
whilst disks are identified by rotational signatures found in both velocity and velocity dispersion maps.
A growing number of massive disks have been found with IFS at $z>2$ which are well fit by disk models and which have Toomre parameters of $Q>1$ indicating stable rotationally supported disks, lending observational support to theories of rapid accretion of cold gas and secular disk evolution at high redshift \citep{2008ApJ...687...59G, 2009A&A...504..789E, 2010MNRAS.402.2291L}.

However, classifying galaxies in these two distinct groups is an over-simplification. Many of the samples present a variety of kinematic types including dispersion-dominated systems, stable disks, unstable disks, minor mergers, mergers, and merger remnants. \cite{2009ApJ...697.2057L} present a sample of lower mass ($M_{*,avg}=2\times10^{10}$~\Msun) galaxies with kinematics dominated by dispersion. Their results can be interpreted as merger remnants, however they also caution that this is statistically unlikely and find that the kinematic classifications are a crude over-simplification. They argue that their galaxies are observed at a range of times during their formation, and that their kinematic signatures are a result of how they acquire their gas.
Despite the different kinematic interpretations, all high-redshift IFS samples show high dispersions ($\sigma \ga$ 90 km s$^{-1}$) with many (20$-$100\%) systems showing no ordered rotation making classifications uncertain  (e.g. \citealt{2008ApJ...687...59G, 2009ApJ...706.1364F, 2009ApJ...697.2057L,2009ApJ...699..421W}).  

With observation times ranging from 2000s -- 40000s, it is only possible to follow up a small fraction of the brightest galaxies at these redshifts. To supplement high-redshift kinematic results other studies target lensed galaxies magnified in both size and flux, allowing the study of sub-L$_*$ galaxies \citep{Jones:2010uf}. Using this strategy, IFUs have probed all the way out to $z=4.9$ with NIFS on Gemini \citep{Swinbank:2009tw} and have targeted strongly lensed galaxies at $z=1.4-3.1$ with OSIRIS on Keck \citep{2008Natur.455..775S,Jones:2010uf,2011ApJ...732L..14Y}. With galaxy magnification individual clumps have been resolved with diameters of $\sim$500 pc which will be essential in testing the models of clumpy disk formation \citep{Jones:2010uf,Swinbank:2009tw,2010Natur.464..733S}. However, more kinematically resolved clumps are needed than are available from current studies ($\sim 20$ at $z>1$) in order to rigorously compare with models.

Despite great advantages, lensed studies require strong lenses with confirmed redshifts which can be observationally expensive in addition to requiring reliable lensing models and reconstruction. Furthermore, rather than bridge the gap between the $z\sim2-3$ results and local galaxies, the majority of current lensing studies push out to higher redshifts.

In order to correctly interpret IFS data at high redshift it is critical to utilize low-redshift IFS results. By comparing directly to these data and by artificially redshifting them out to $z>1$ a better understanding of different kinematic signatures in the high-redshift data can be reached.  Recent efforts by \citet{2010MNRAS.401.2113E} have supplied a sizable sample of FabryÐPerot 3D data of \halpha emission in 153 nearby isolated galaxies. Disk models are fit to the local population and to the same galaxies artificially redshifted to $z=1.7$, providing guidance in the best disk models to use to fit to high-redshift kinematic results. \citet{Green:2010fk} and \citet{2011A&A...527A..60R} also provide local comparison samples for \halpha velocity dispersion and morphology. Properties observed locally at parsec to kiloparsec resolution are crucial for the interpretation of high-redshift IFU observations.

Connecting the $z<0.2$ and $z>1.5$ results is a relative dearth of IFS studies closer to $z\sim1$. At this redshift, galaxies are at the tail end of the epoch of mass assembly but at more accessible distances for current instrumentation (where $(1 + z)^{4}$ surface brightness dimming is less severe). Yet suitable galaxies for such studies are rare as once the main epoch of massive galaxy formation is over at $z\sim2$, the abundance of galaxies with high star formation rates falls rapidly. They are difficult to select with the Lyman-break technique at $z\sim1$ as the drop in emission blueward of 912\AA~ cannot be detected with the low sensitivity and resolution of ultraviolet (UV) instruments. These galaxies can, however, be found in very large high-redshift surveys, such as the WiggleZ Dark Energy Survey.  The WiggleZ survey contains a rare population of UV-luminous galaxies out to $z=1.5$ with star formation rates (from nebular emission) of $\sim$100 M$_{\odot}$ yr$^{-1}$. These galaxies may be the link needed to bridge the gap between observations of local luminous galaxies, and LBGs and SMGs found at higher redshift.
In this paper, we present integral-field spectroscopy of the \halpha emission line in 13 star-forming galaxies from the WiggleZ Dark Energy Survey at $z\sim1.3$.

\begin{figure}
\includegraphics[scale=0.55]{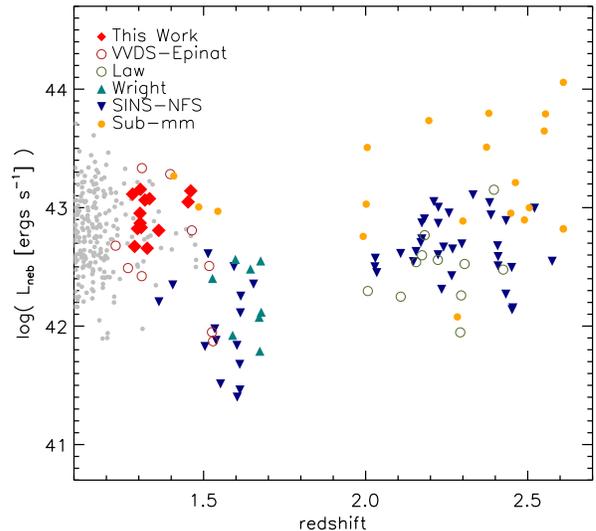}
\caption{
Nebular luminosity as a function of redshift for the largest high-redshift IFS surveys with the work presented here shown by the red diamonds. The gray points show the \OII~luminosity of other WiggleZ galaxies considered for this sample. The red circles show the sample from \citet{2009A&A...504..789E} taken from the VVDS survey and observed with SINFONI on the VLT. The green circles show the sample from \citet{2009ApJ...697.2057L} observed with OSIRIS on Keck. The blue inverted triangles show the SINS sample from \citet{2009ApJ...706.1364F} observed with SINFONI, and the green triangles show the sample from \citet{2009ApJ...699..421W} observed with OSIRIS. The orange points are SMGs with measured \halpha taken from \citet{2004ApJ...617...64S} and \citet{2006ApJ...651..713T}.}
\label{fig.litcomp}
\end{figure}

In this paper, we present 13 galaxies and their physical properties. In Section 2 we describe our sample selection, observational technique, and data reduction methods. In Section 3 we discuss the properties of the galaxies, in particular we describe the morphological and kinematic properties. In Section 3.6 and 3.7 we discuss the methods of disk fitting and surface brightness profile fitting analysis. The individual and global properties of the 13 systems described in Section 3 are analyzed in Section 4 in the context of theories of galaxy formation at high redshift and the implications for the mechanism triggering massive rates of star formation. We conclude with a summary of our results and interpretations in Section 5, giving a brief list of the key points presented in this paper. The properties of the 13 individual galaxies are described individually in the Appendix.

A standard $\Lambda$CDM cosmology of $\Omega_{\mathrm{m}} = 0.3$, $\Omega_{\Lambda} = 0.7$, $h = 0.7$ is adopted throughout this paper (e.g. \citealt{2011ApJS..192...18K}). In this cosmology, at redshift $z=1.3$, 1 arcsec corresponds to 8.38 kpc.

\section{Observations}
       
\subsection{Target Selection} 
Our sample of 13 galaxies was selected from the WiggleZ Dark Energy Survey \citep{Drinkwater:2010bx}.  WiggleZ is a UV-selected survey of 238,770 unique emission line galaxies in the redshift range $0 < z < 1.5$, measured over 245 nights with the AAOmega spectrograph \citep{2006SPIE.6269E..14S} on the 3.9-m Anglo-Australian Telescope (AAT).  The primary aim of the survey is to precisely measure the scale of baryon acoustic oscillations imprinted on the spatial distribution of these galaxies at look-back times of 4--8 Gyrs \citep{2011MNRAS.tmp}. The survey samples a total volume of $\sim$1 Gpc$^{3}$ over 816 deg$^{2}$ of sky across seven fields, with an average target density of 350 galaxies per deg$^{2}$. Photometric data used in target selection for the WiggleZ survey are from the the Galaxy Evolution Explorer (GALEX) satellite \citep{2005ApJ...619L...1M}, the fourth data release of the Sloan Digital Sky Survey (SDSS; \citealt{2006ApJS..162...38A}) in the northern galactic region, and the Canada-France-Hawaii Telescope Second Red-sequence Cluster Survey (RCS2;  \cite{2011AJ....141...94G}) in the southern galactic region. 

Whilst the WiggleZ galaxy redshift distribution peaks at $z\sim0.6$ a long thin tail of high-redshift galaxies reaches to $z=1.5$.  These high-redshift galaxies are identified by strong \OII~emission lines, with the \OII~doublet marginally resolved, and have star formation rates of $>50$\sfrunits. 

Galaxies selected for this study were drawn from the WiggleZ survey subject to a variety of criteria. Primary selection required (1) redshifts of $1.2<z<1.5$, where the \OII~emission line falls between OH sky residuals at $\sim$8000--10000 \AA, and (2) \OII~emission exceeding $50 \times 10^{-15}$ ergs s$^{-1}$  cm$^{-2}$. We found $\sim$300 targets in the WiggleZ survey that satisfied the aforementioned criteria. To reduce contamination from active galactic nuclei (AGN), starbursts were preferentially selected by  (3) omitting galaxies with [Mg\hspace{.03cm}II] emission visible in the WiggleZ spectrum which may be indicative of AGN activity \citep{1979ApJ...232..659G}. 
With the additional criteria the number of targets was reduced to $\sim$100 galaxies. Finally, for IFU observations with laser guide star adaptive optics (LGSAO), a suitable tip-tilt star is required to correct aberrations in the wavefront.  We require a star with apparent r-band magnitude $m_{\mathrm{r}} \la 17$ with an angular separation $<60$ arcsec from the science target for optimal acquisition. This last criteria reduced our final target sample to $\sim$50 targets.   

\begin{figure*}
\includegraphics[scale=1.01]{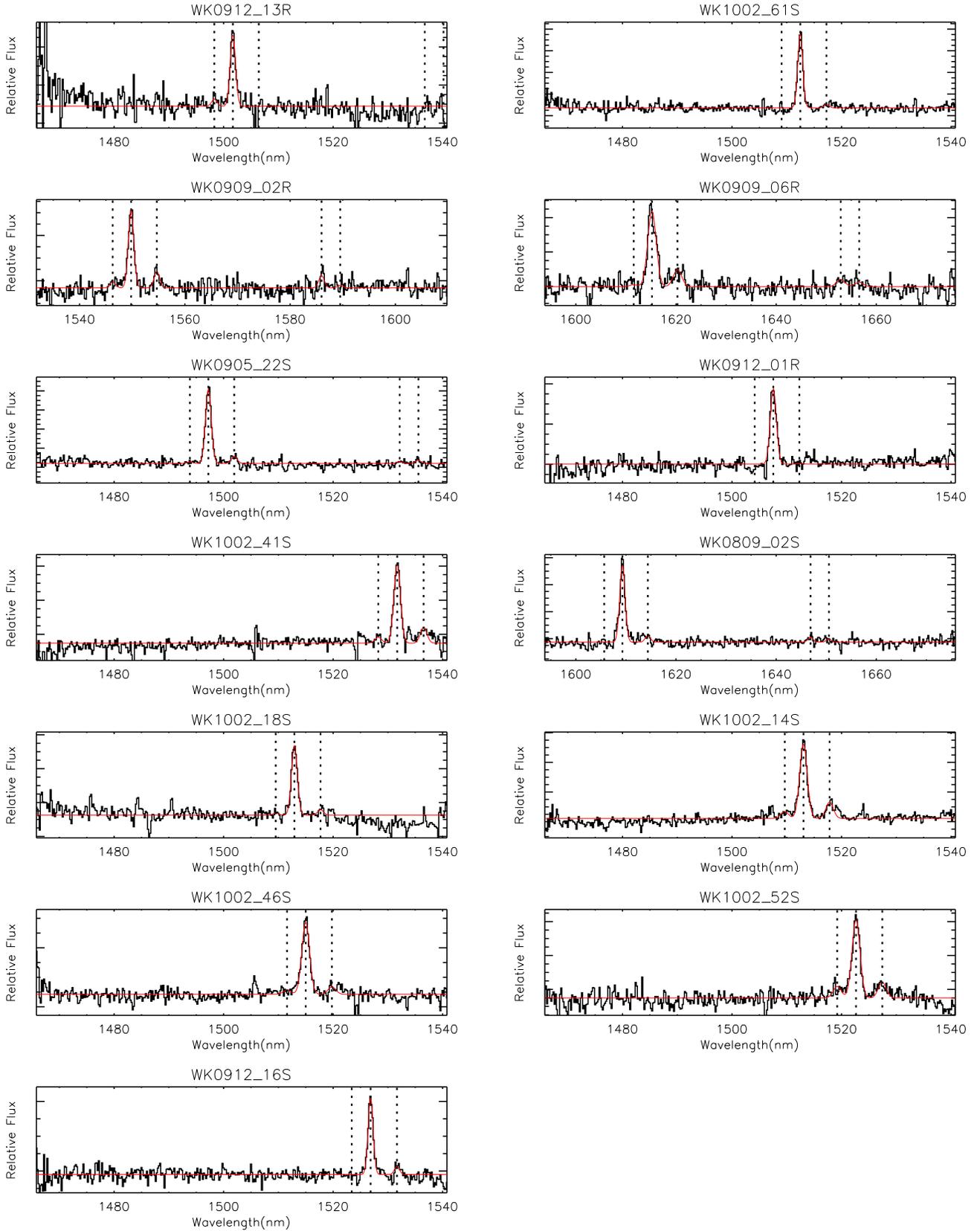}
\caption{
OSIRIS integrated spectra of each galaxy over all pixels satisfying the criteria outlined in Section 2. The vertical dotted lines denote the position of the redshifted locations of nebular emission lines: \halphans, \NII$\lambda$$\lambda$6548, 6584, and \SII$\lambda$$\lambda$6716, 6731 at the systematic redshift. The solid red line is the best-fitting single Gaussian fits to each detected emission line.
\label{fig.intspec}
}
\end{figure*}

\setcounter{figure}{2}
\begin{figure*}
\includegraphics[scale=0.7]{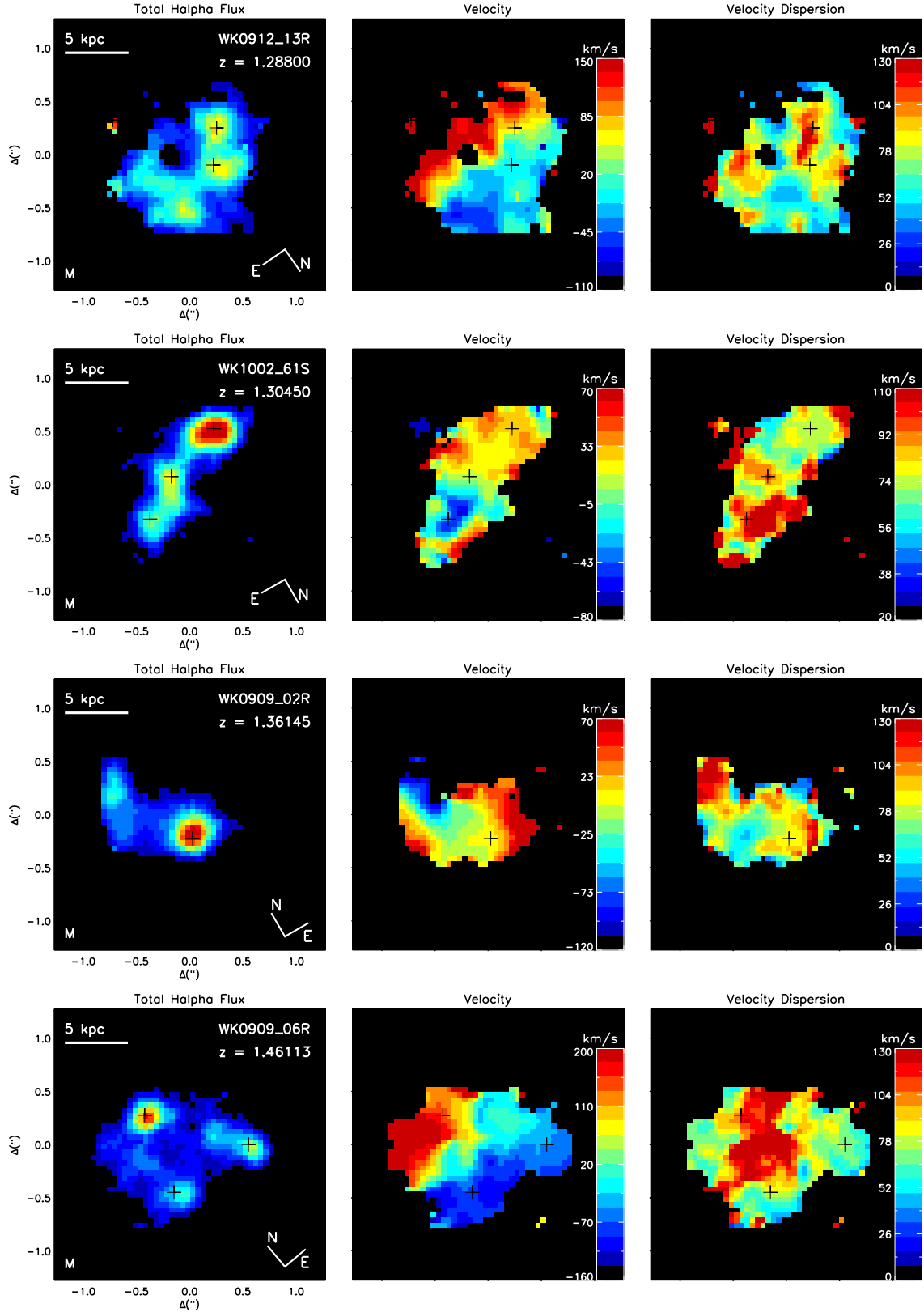}
\caption{
OSIRIS \halpha kinematics. {\it Left:} \halpha flux map, {\it middle:} velocity map from the systemic velocity, {\it right:} velocity dispersion map. The black crosses mark the peak(s) in \halpha flux. Velocity shifts from the systemic redshift and velocity dispersion are measured in km s$^{-1}$.  Compass at the bottom right of the left panels show the true orientation. Letters at the bottom left of the left panels represent morphology classifications; M=Multiple emission, E=Extended emission, S=Single emission. Objects are presented in the order based on the morphological classification described in Section 3.1. 
\label{fig.kinematics}
}
\end{figure*}

\setcounter{figure}{2}
\begin{figure*}
\includegraphics[scale=0.7]{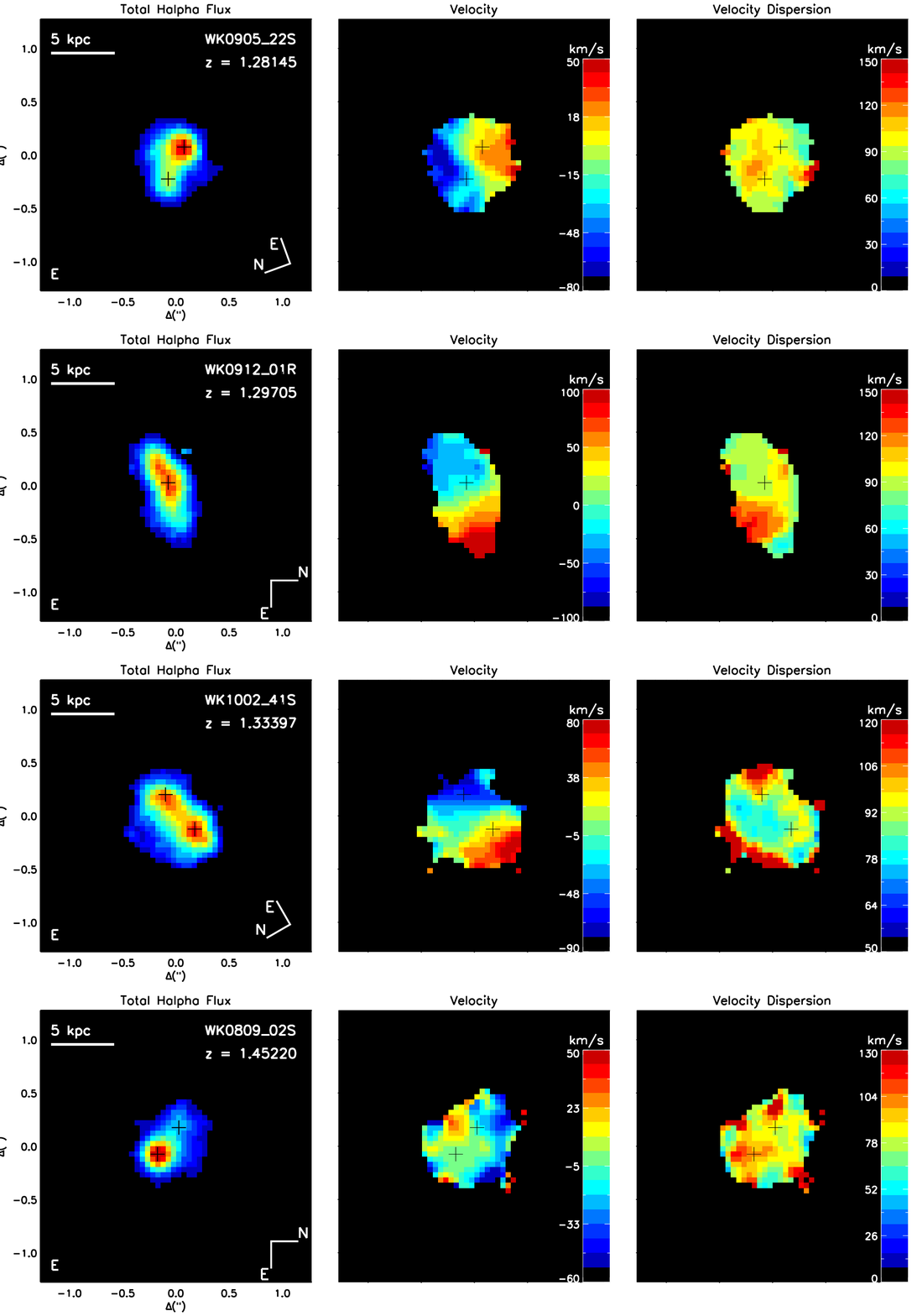}
\caption[justification=justified,singlelinecheck=false]{{\it cont.}
\label{fig.kinematics}
}
\end{figure*}

\setcounter{figure}{2}
\begin{figure*}
\includegraphics[scale=0.7]{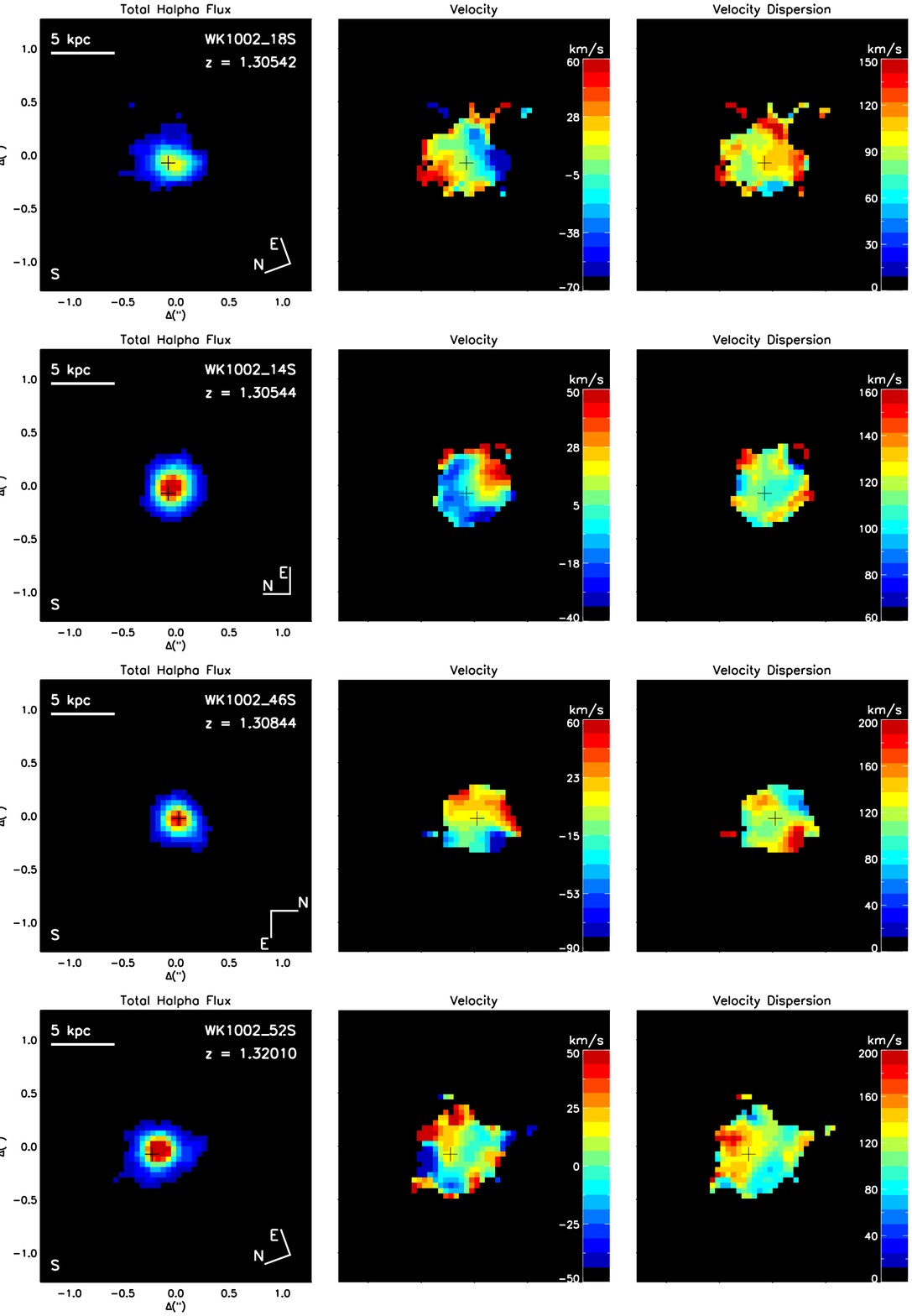}
\caption{{\it cont.}
\label{fig.kinematics}
}
\end{figure*}

\setcounter{figure}{2}
\begin{figure*}
\includegraphics[scale=0.7]{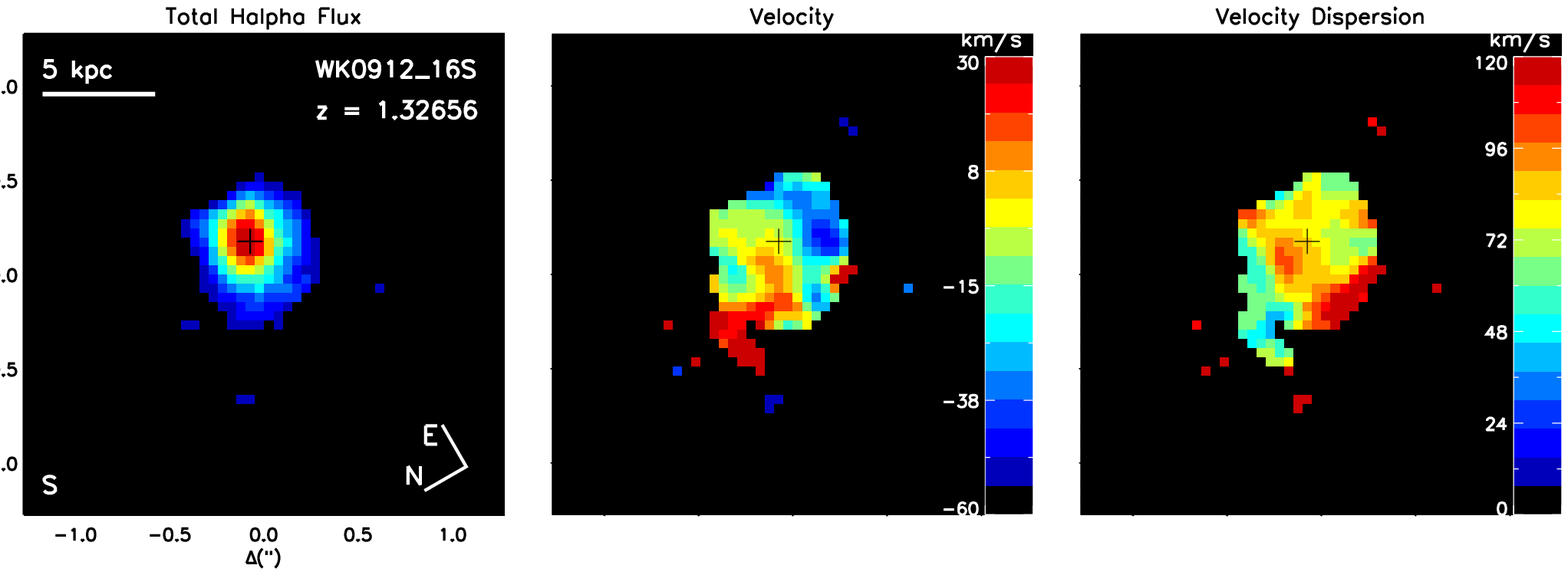}
\caption{{\it cont.}
\label{fig.kinematics}
}
\end{figure*}

\begin{table*}
\begin{minipage}{\textwidth}
\caption{Emission Line Properties}
\begin{tabular*}{\textwidth}{@{\extracolsep{\fill}}llcrcccc}
\hline
{WiggleZ ID} & {ID	} & 
{$z_{\mathrm{sys}}^{a}$} & 
{$f_{\mathrm{H\alpha}}^{b}$	} & 
{$f_{\mathrm{\NII}}^{b}$	} & 
{$L_{\mathrm{H\alpha}}$} 
& {$12+\mathrm{log}$(O/H) } & {} \\
{} & {} & {} & {} & {} & {(10$^{41}$ ergs~s$^{-1}$)} & {} & {} \\
\hline
R03J032450240$-$13550943 & WK0912\_13R &       1.2873 & 48    $\pm$ 16.1 &    $<$8.4      & 46.8 $\pm$ 13.2   &    $<$8.47 \\
S15J145355248$-$00320351 & WK1002\_61S &       1.3039 & 81.5  $\pm$ 5.3  &    $<$9.4      & 81.9 $\pm$ 13.7   &    $<$8.37\\
R01J005822757$-$03034040 & WK0909\_02R &       1.3616 & 41.5  $\pm$ 7.9  & 9.3 $\pm$ 4.4  & 46.4 $\pm$ 15.2   & 8.53 $\pm$ 0.29 \\
R03J032206214$-$15443471 & WK0909\_06R &       1.4602 & 94.9  $\pm$ 6.7  & 22.8 $\pm$ 3.7 & 126.2 $\pm$ 18.1  & 8.55 $\pm$ 0.1 \\

S15J144102444+05480354   & WK0905\_22S &       1.2810 & 105.4 $\pm$ 4.3  & 10.8 $\pm$ 2.3 & 101.4 $\pm$ 13.1  & 8.34 $\pm$ 0.12 \\
R00J232217805$-$05473356 & WK0912\_01R &       1.2960 & 57.3  $\pm$ 7.5  &    $<$6.4      & 56.7 $\pm$ 13.5   &    $<$8.36 \\
S09J091517481+00033557   & WK1002\_41S &       1.3330 & 64.9  $\pm$ 4.3  & 13.7 $\pm$ 2.4 & 68.9 $\pm$ 14.4   & 8.52 $\pm$ 0.1 \\
S00J233338383$-$01040629 & WK0809\_02S &       1.4521 & 74.7  $\pm$ 13.6 & 6.3 $\pm$ 7.3  & 98.0 $\pm$ 17.8   & 8.29 $\pm$ 0.68 \\

S15J142538641+00483135   & WK1002\_18S &       1.3044 & 58.0  $\pm$ 4.8  &    $<$7.6      & 58.4 $\pm$ 13.7   &    $<$8.4 \\
S11J101757445$-$00244002 & WK1002\_14S &       1.3048 & 92.8  $\pm$ 4.3  & 20.6 $\pm$ 2.4 & 93.5 $\pm$ 13.7   & 8.53 $\pm$ 0.07 \\
S09J090933680+01074587   & WK1003\_46S &       1.3074 & 50.0  $\pm$ 4.5  &    $<$5.6      & 50.6 $\pm$ 13.8   &    $<$8.36 \\
S11J110405504+01185565   & WK1003\_52S &       1.3191 & 86.1  $\pm$ 6.0  & 20.4 $\pm$ 3.4 & 89.0 $\pm$ 14.1   & 8.54 $\pm$ 0.1 \\
S09J090312056$-$00273273 & WK0912\_16S &       1.3260 & 38.6  $\pm$ 4.4  & 7.5 $\pm$ 2.4  & 40.5 $\pm$ 14.3   & 8.49 $\pm$ 0.19 \\
\hline
\label{flux.table}
\end{tabular*}\\
$^{a}$Systematic redshift measured from Gaussian fit to \halpha emission in integrated spectrum.\\
$^{b}$Emission line flux in units of 10$^{-17}$ ergs~s$^{-1}$~cm$^{-2}$. 
\end{minipage}
\end{table*}

\subsection{OSIRIS Observations}
Out of the possible targets, spanning across seven fields, we observed 17 star-forming galaxies with OSIRIS (OH Suppressing InfraRed Imaging Spectrograph; \citealt{2006NewAR..50..362L}) with the Keck II LGSAO system \citep{2006PASP..118..297W, 2006PASP..118..310V}. These 17 targets were prioritized for having the brightest tip-tilt stars with shortest separation, and for which \OII~and \halpha fall fortuitously in clean sky regions. 
OSIRIS is a lenslet array spectrograph with a $2048 \times 2048$ Hawaii-2 detector and spectral resolution $R\sim3600$. Observations were performed in September 2008, May, September, and December 2009, and March 2010. The filters Hn1, Hn2, and Hn3 were used with central wavelengths of 1.500, 1.569, 1.635 $\mu$m respectively in the 0.05$''$ pixel scale.  The FWHM of the tip-tilt stars ranged from one and a half to two pixels, 0.062-0.1$''$, with an average Strehl of 30\%  estimated from the tip-tilt stars. Seeing ranged from 0.4$''$ to 1.4$''$. The galaxies observed on these nights are listed in Table~\ref{osirisobs.table}.

The standard observing procedure was as follows. We first acquired the tip-tilt star and applied the optimal position angle of OSIRIS to position the star within the un-vignetted field-of-view of the LGSAO system. Short, $60$s, integrations were taken on the star for calculations of the PSF and to centre the star in the field. A blind offset was then applied to the telescope to acquire the target. The WiggleZ survey selected targets from SDSS and RCS2 which follow different astrometry systems \citep{Drinkwater:2010bx}. In calculating offsets from the reference star to the target, 2MASS \citep{2006AJ....131.1163S} and USNO-B \citep{2003AJ....125..984M} coordinates were used for SDSS and RCS2 galaxies respectively. The typical science observing sequence was to expose on target for 900s in each of two nod positions, A and B,  separated by 1.6$''$ on the detector. If \halpha emission was detected in the difference of the first two exposures, we continued to expose on the galaxy for $\sim1-2$ hrs as necessary to achieve a high signal-to-noise detection. Multiple A and B positions were dithered by $0.05''$ around the base positions in each exposure to remove bad pixels and cosmic ray contamination. Of the 17 observed targets, \halpha was detected in the spectra of 15 targets. The two non-detections were a result of the \OIII$\lambda$5007 emission line being mis-identified as \OII$\lambda$3727 in the WiggleZ spectrum. Of the 15 detected targets, observations for two galaxies are unusable for scientific reduction due to poor AO correction as a result of stray light on the low bandwidth wavefront sensor. The \halpha luminosities of the final 13 galaxies are shown in Figure~\ref{fig.litcomp} with \OII~luminosities of all possible targets indicated by the small gray filled circles.

Data reduction was completed with the OSIRIS data reduction pipeline versions 2.2 and 2.3 and custom IDL routines developed for faint emission-line spectra. The pipeline removes crosstalk, detector glitches, and cosmic rays before it mosaics individual exposures and assembles a reduced data cube with two spatial and one wavelength dimension. First-order sky subtraction was achieved by the spatial ABBA nodding on the sky. Further sky subtraction was applied using custom IDL routines which employ the sky subtraction methods of \cite{2007MNRAS.375.1099D}. Galaxies were flux calibrated using telluric standard stars observed throughout the night. On average three telluric standards were observed each night. 


\begin{table}
\begin{minipage}{\textwidth}
\caption{IRIS2 Observations of WiggleZ Galaxies}
\begin{tabular}{@{\extracolsep{\fill}}lccccccc}
\hline
{ID	} & 
{Obs.} & 
{J Exp.} & 
{K Exp.} & 
{J$^{a}$} & 
{K$^{a}$} \\
{} & 
{Date} & 
{time(s)	} & 
{time(s)} &
 {} & 
 {} \\
\hline
WK0912\_13R &   2010/01 & 1350  &	  4050	&  20.71   &  20.13 \\
WK0909\_06R &  	2010/01	& 2700  &	  3360	&  20.77   &  19.98 \\

WK0905\_22S &   2009/05	& 1350  &	  1350	&  21.36   &  20.52 \\         
WK1002\_41S &   2010/01 & 1350  &	  3600	&  21.30   &  20.57 \\
WK0809\_02S &   2010/07	& 1950  &     $\ldots$	& 20.66 & $\ldots$ \\ 

WK1002\_18S &   2009/05	& 1350  &	  2700	&  20.84   &  21.02 \\
WK1002\_14S &   2009/05	& 1350  &	  1350	&  20.69   &  20.20 \\
WK1002\_52S &   2010/01	& 1350  &	  3300	&  19.94   &  18.92 \\
WK0912\_16S &  	2010/01	& 1260  &     4050	&  21.79   &  21.93 \\          
\hline
\multicolumn{6}{l}{$^{a}$Apparent AB magnitudes. The average error on J and K}\\ 
\multicolumn{6}{l}{magnitudes are 0.11}
\label{iris2.table}
\end{tabular}
\end{minipage}
\end{table}

\subsection{Kinematic Mapping}
In order to create 2-D spatial and kinematic maps of the \halpha emission line, the sky-subtracted data cube was convolved with a Gaussian kernel with $FWHM=150$ mas (200 mas in cases of low signal) to achieve a higher signal-to-noise ratio in each spaxel. A Gaussian profile was then fitted to the \halpha emission in each spaxel of the cube with the wavelength, intensity and width as free parameters. Errors were calculated by adding in quadrature an estimation of the sky noise from the variance of the spectrum offset from the emission lines with the Poisson error of the photon counting statistics in the emission lines.  Spaxels were masked from the \halpha emission map and kinematic maps if the peak of the fitted Gaussian was less than one standard deviation above the median of the spectrum (corresponding to minimum signal-to-noise of $\sim$2), or the width of the Gaussian was less than the resolving power of the instrument. For spaxels surviving these cuts a flux was calculated from the fitted Gaussian. The \halpha flux images represent this array of fluxes. The integrated spectra, shown in Figure~\ref{fig.intspec}, were created by summing all unmasked spaxels in the spatial directions. A Gaussian was fitted to the observed-frame integrated spectrum to obtain the total \halpha flux and the systematic redshift, $z_{\mathrm{sys}}$ given in Table~\ref{flux.table}. A velocity map was constructed for each galaxy using the velocity shifts of the emission line in each spaxel relative to $z_{\mathrm{sys}}$. The velocity dispersion was measured in each spaxel from the r.m.s.\ of the fitted Gaussian line profile corrected for instrumental broadening. All kinematic maps are shown in Figure~\ref{fig.kinematics}.

\subsection{NIRSPEC Observations \& Data Reduction}
When the Keck LGSAO system was unavailable due to bad weather, technical faults or space events, observations of available targets were taken with the near-infrared echelle spectrograph (NIRSPEC; \citealt{1998SPIE.3354..566M}) to measure the $H\beta$ and \OIII~emission line ratio. The ratios of emission lines from these observations, obtained for $WK0905\_22S$, $WK1002\_18S$, $WK1002\_52S$, $WK1002\_61S$, are used to constrain their AGN contamination as discussed in Section 3.4.

The NIRSPEC detector is a 1024$\times$1024 ALADDIN-3 InSb detector with spectral resolution $R\sim2000$. All observations were carried out in February 2010 with the 4-pixel slit (42$\times$0.76$''$) in the NIRSPEC-2 filter (1.089$-$1.293 $\mu$m). Seeing in K was 0.8$''$ on average.

The typical observing sequence was to acquire a guide star, typically the star already selected for use with OSIRIS LGSAO system. Short ($\sim$10s) images were taken with the NIRSPEC guider to center the star at the central position on the slit. Once centered a blind offset was applied to the star position. Longer images were taken ($2\times20$s) at the target position. If the target was observed in these images it was centered on the slit by nudging the slit left and right in single pixel intervals. 
Once centered spectroscopic object exposures were taken with typical sequence of $2\times600$s in each of two nod positions, A and B in the sequence ABBA. Occasional shifts of a few pixels were needed to keep the object on the slit throughout the object exposures. No continuum was detected and the ratio of \OIII$\lambda5007$/H$\beta$$\lambda4861$ was calculated from the raw data.

\begin{table*}
\begin{minipage}{\textwidth}
\caption{\halpha Kinematic \& Morphological Properties}
\begin{tabular*}{\textwidth}{@{\extracolsep{\fill}}lcccccc}
\hline
{ID	} & {$v_{\mathrm{shear}}^{a}$} & 
{$\sigma_{\mathrm{mean}}^{b}$	} & 
{$\sigma_{\mathrm{net}}^{c}$	} & 
{ $v_{\mathrm{shear}}$/ $\sigma_{\mathrm{mean}}$} &
{$r_{1/2}^{d}$} & 
{ Class$^{e}$    } \\
{} & 
{(km s$^{-1}$)	} & 
{(km s$^{-1}$)	} & 
{(km s$^{-1}$)	} & 
{} & 
{(kpc)} & 
{            } \\
\hline
WK0912\_13R & 130.5  $\pm$ 11.0 & 81.6 $\pm$ 27.7 & 104.3 $\pm$ 27.1 & 1.6 $\pm$ 0.6 & 3.8 & M\\
WK1002\_61S &  71.5  $\pm$ 10.7 & 85.1 $\pm$ 20.3 & 88.4 $\pm$ 4.4 & 0.9 $\pm$ 0.2      & 3.8 & M\\
WK0909\_02R & 136.4  $\pm$ 6.1 & 87.9 $\pm$ 21.9 & 103.7 $\pm$ 15.5 & 1 $\pm$ 0.3      & 2.6 & M\\
WK0909\_06R & 160.8  $\pm$ 2.4 & 92.6 $\pm$ 27.8 & 153.3 $\pm$ 8.4 & 1.6 $\pm$ 0.5     & 4.2 & M\\
              
WK0905\_22S &  55.5  $\pm$ 6.5 & 98.2 $\pm$ 10.9 & 107.4 $\pm$ 3.3 & 0.6 $\pm$ 0.1      & 2.3 & E\\
WK0912\_01R &  99.4  $\pm$ 7.1 & 98.6 $\pm$ 19.6 & 112.1 $\pm$ 11.0 & 1.3 $\pm$ 0.3  & 2.8 & E\\
WK1002\_41S &  80.2  $\pm$ 5.4 & 91.4 $\pm$ 13.0 & 116.3 $\pm$ 5.9 & 1.7 $\pm$ 0.3       & 3.1 & E\\
WK0809\_02S &  48.5  $\pm$ 9.4 & 92.7 $\pm$ 16.6 & 97.6 $\pm$ 13.8 & 0.4 $\pm$ 0.1      & 2.1 & E\\
              
WK1002\_18S &  81.1  $\pm$ 17.5 & 102.1 $\pm$ 20.0 & 106.6 $\pm$ 6.8 & 0.7 $\pm$ 0.2  & 2.1 & S\\
WK1002\_14S &  48.4  $\pm$ 9.5 & 113.7 $\pm$ 8.8 & 133.3 $\pm$ 4.8 & 0.5 $\pm$ 0.09     & 2.1 & S\\
WK1003\_46S &  65.6  $\pm$ 16.7 & 124.8 $\pm$ 24.9 & 138.6 $\pm$ 9.6 & 0.5 $\pm$ 0.2    & 1.8 & S\\
WK1003\_52S &  73.7  $\pm$ 18.2 & 125.4 $\pm$ 20.9 & 145.4 $\pm$ 8.1 & 0.7 $\pm$ 0.2  & 1.8 & S\\
WK0912\_16S &  54.9  $\pm$ 12.9 & 85.1 $\pm$ 18.6 & 104.4 $\pm$ 9.1 & 0.7 $\pm$ 0.2    & 2.5 & S\\
\hline
\end{tabular*}\\
{$^{a}$Velocity shear measured across the \halpha velocity map as described in Section 3.2.}\\
{$^{b}$Flux-weighted mean velocity dispersion calculated from a Gaussian fit to the spectra of individual spaxels of the OSIRIS \halpha detection.}\\
{$^{c}$Velocity dispersion calculated from a Gaussian fit to the integrated spectrum of the OSIRIS \halpha detection.}\\
{$^{d}$Half-light radius of OSIRIS \halpha emission in circular aperture.}\\
{$^{e}$Morphological classification; E:Extended emission, M:Multiple emission, S:Single emission.}\\ 
\label{kinematics.table}
\end{minipage}
\end{table*}

\subsection{IRIS2 Observations}
Near-infrared imaging was obtained for 9 out of 13 galaxies in the sample with the Infrared Imager and Spectrograph (IRIS2,  \citealt{2004SPIE.5492..998T}) on the Anglo-Australian Telescope (AAT). $J$ and $Ks$ bands, with central wavelengths $\lambda_{c}=1.245~\mu$m and $\lambda_{c}=2.1444~\mu$m respectively, were chosen in order to image the older stellar population of each galaxy to obtain robust estimates of stellar masses.  The IRIS2 detector is a $1024 \times 1024$ Rockwell HAWAII-1 HgCdTe infrared detector, with a scale of 0.4487 arcsec pixel$^{-1}$ and field of view of 7.7 arcmin$^{2}$. Observations were carried out during May 2009, January 2010, and July 2010. Seeing in $Ks$ ranged between 0.9$''$-2.5$''$,  0.9$''$-1.6$''$, and 1.3$''$-2.0$''$, respectively. 

The typical $J$ exposure sequence was a series of $15$ second exposures totaling $1350$ seconds on the target.  The typical $Ks$ exposure sequence was a series of $10$ seconds exposures totaling $1350$ seconds on target. Most galaxies required 2 or 3 sequences in $Ks$ in order to obtain a detection with a signal-to-noise ratio greater than 5. Due to the time-sensitive variation of the infrared night sky, we used a dithering strategy of successive observations with 9 dithers per sequence with a maximum dither length of 20 arcsec. 

Data reduction and mosaicking was carried out by data reduction pipeline \oracdr \citep{2003ASPC..295..237C}. Each image was flattened with the other images closest in time and position in the dither pattern. The dithering script averaged out bad pixels on the detector and applied a nearest neighbor smoothing. 

Magnitudes were calculated in the AB system using aperture photometry with a circular aperture of $4''$ with a concentric sky annulus. A zero point was derived for each reduced mosaicked image by comparing the measured aperture counts with known 2MASS magnitudes for several stars in the image. The zero point was then used to photometrically calibrate the image. On average, four uncrowded central stars with $K$-band magnitudes of $K\sim15$ were selected to derive the zero point. If the standard deviation of the zero points from these stars was greater than 0.3, the stars producing the most deviant zero points were replaced with other stars in the field. Errors were calculated from the variance in the sky aperture. The sky background was calculated using a clipped mean. The IRIS2 system most closely resembles the Mauna Kea Observatory (MKO) system. Equations~\ref{eq.photomj} and \ref{eq.photomk} were used to  transform between the MKO and 2MASS systems,

\vspace{-0.3cm}
\begin{eqnarray}
J_{MKO} &=& -0.03 - 0.03 \times (J-Ks) + J 
\label{eq.photomj}\\
K_{MKO} &= &-0.01 - 0.01 \times (J-Ks) + K
\label{eq.photomk}
\end{eqnarray}

where $J$ and $Ks$ are given in the 2MASS system and $J_{MKO}$ and $K_{MKO}$ are given in the Mauna Kea Observatory system used by IRIS2\footnote[1]{http://www.aao.gov.au/iris2/}. Derived values of $J$ and $K$-band magnitudes are given in Table~\ref{iris2.table}.

\subsection{Stellar Mass Fitting}
Stellar masses and star formation rates were calculated using an spectral energy distribution (SED) fitting routine \citep{2004Natur.430..181G} which employs the Baldry \& Glazebrook initial mass function (IMF) (BG03; \citealt{2003ApJ...593..258B}). The BG03 IMF fits local cosmic luminosity densities well, with a high-mass slope similar to the classical Salpeter IMF but with a more realistic break at 0.5\Msun. \cite{2006ApJ...651..142H} have shown that the BG03 IMF gives a better agreement between cosmic star-formation histories and mass assembly. 

Infrared {\it JK} data from IRIS2 was combined with multiband photometry available in WiggleZ fields, including {\it ugriz} from SDSS, {\it grz} from the RCS2 Survey, and {\it NUV} from GALEX. Two-component modeling was adopted (using PEGASE.2 from \cite{1997A&A...326..950F} to calculate spectra) to assess the possible biases from starbursts on the calculated masses and to allow a range of dust extinction (0 $\leq A_V \leq$ 2 mag) and metallicity (0.0004 $\leq Z \leq$ 0.02). The primary component was modeled using SFR $\propto$ exp(-t/$\tau$) with $\tau=$0.1, 0.2, 0.5, 1, 2, 4, 8, \& 500 Gyr (the first approximates an instantaneous starburst and the last a constant SFR). The full distribution function of allowed masses was calculated by Monte-Carlo re-sampling of the photometric errors.  The final masses and error bars, given in Table~\ref{stellar.table}, represent the mean and standard deviation of this full distribution function. In the cases where observed UV, optical and NIR magnitudes were available the errors in the derived stellar masses are on average $\delta$[log(M$_{*}$[\Msun])] = 0.15. When NIR magnitudes were not available, the errors average $\delta$[log(M$_{*}$[\Msun])]= 0.62. 

We do not use the models of \cite{2006ApJ...652...85M} as \cite{2010ApJ...722L..64K} show a low contribution from TP-AGB stars for post-starburst galaxies at high redshift. For more information on the SED fitting see \cite{2004Natur.430..181G}. 

\begin{table*}
\begin{minipage}{\textwidth}
\caption{Stellar Properties}
\begin{tabular*}{\textwidth}{@{\extracolsep{\fill}}lrcrcccc}
\hline
{ID	} & 
{log$[ M_{*} (\Msun)] $} & 
{$A_V$} &
{SFR$_{\mathrm{H\alpha}}$ } & 
{SFR$_{\mathrm{\OII}}^{a}$ } & 
{SFR$_{\mathrm{SED}}$ } & 
{SSFR} & 
{            } \\
{} & 
{} & 
{(mag)} &
{(\sfrunits)} & 
{(\sfrunits)} & 
{(\sfrunits)} & 
{($10^{-9} \mathrm{yr}^{-1}$)} & 
{            } \\
\hline
WK0912\_13R &   10.7  $\pm$  0.2 & 1.1 $\pm$ 0.3 &   20.3 $\pm$ 6.8 &     43.3  $\pm$ 0.9  &        116  $\pm$ 40.4 &   1.8 \\
WK1002\_61S &   10.7  $\pm$  0.7  & 0.2 $\pm$ 0.2 &   35.6 $\pm$ 2.3 &     69.0  $\pm$ 2.8  &       25.6 $\pm$ 13   &   2.9 \\
WK0909\_02R &   10.3  $\pm$  0.6 & 0.7 $\pm$ 0.4 &   20.2 $\pm$ 3.8 &     33.1  $\pm$ 0.8  &        57.4 $\pm$ 41.5 &   4.6 \\
WK0909\_06R &   11.0  $\pm$  0.2 & 0.8 $\pm$ 0.3 &   54.8 $\pm$ 3.9 &     120.3 $\pm$ 14.9 &     91.8 $\pm$ 48.7 &   2.09 \\
                            
WK0905\_22S &   10.3  $\pm$  0.2 & 0.3 $\pm$ 0.2 &   44.0 $\pm$ 1.8 &     202.4 $\pm$ 23.8 &      44.6 $\pm$ 15.1 &   10 \\
WK0912\_01R &   10.0  $\pm$  0.6 & 0.5 $\pm$ 0.2 &   24.6 $\pm$ 3.2 &     59.0  $\pm$ 3.8  &        40.3 $\pm$ 13   &   11.2 \\
WK1002\_41S &   10.2  $\pm$  0.1 & 0.4 $\pm$ 0.1 &   29.9 $\pm$ 2.0 &     112.8 $\pm$ 5.7  &       60.3 $\pm$ 9.3  &   8.1 \\
WK0809\_02S &   11.1  $\pm$  0.6 & 0.0 $\pm$ 0.1 &   42.5 $\pm$ 7.7 &     29.7  $\pm$ 0.8  &       19.5 $\pm$ 6.9  &   1.5 \\
                            
WK1002\_18S &   10.2  $\pm$  0.1 & 0.5 $\pm$ 0.1 &   25.3 $\pm$ 2.1 &     188.6 $\pm$ 14.2 &      64   $\pm$ 17.4 &   6.09 \\
WK1002\_14S &   10.6  $\pm$  0.2 & 0.3 $\pm$ 0.1 &   40.6 $\pm$ 1.9 &     43.9  $\pm$ 0.7  &       55.5 $\pm$ 16   &   4.5 \\
WK1003\_46S &    9.8  $\pm$  0.6 & 0.2 $\pm$ 0.3 &   22.0 $\pm$ 2.0 &     41.3  $\pm$ 4.8  &        22.4 $\pm$ 17.1 &   12.8 \\
WK1003\_52S &   11.7  $\pm$  0.1 & 0.0 $\pm$ 0.1 &   38.6 $\pm$ 2.7 &     103.7 $\pm$ 16.9 &      15.9 $\pm$ 2.2  &   0.3 \\
WK0912\_16S &    9.8  $\pm$  0.2 & 0.1 $\pm$ 0.1 &   17.6 $\pm$ 2.0 &     64.1  $\pm$ 6.8  &          12.6 $\pm$ 3.2  &   12 \\
\hline
\end{tabular*}\\
{$^{a}$Errors on the \OII~ SFRs are statistical errors only.}\\
\label{stellar.table}
\end{minipage}
\end{table*}

\section{Results}
In Figure~\ref{fig.kinematics} we present \halpha spatial {\it (left panels)} and kinematic maps of our 13 galaxies observed with OSIRIS with spatial extent ranging from $2-12$ kpc. A broad range of morphologies and kinematics are observed with 8/13 galaxies showing some velocity shear and the remaining showing disturbed kinematic morphologies {\it (middle panels)}. The dispersion maps for the entire sample are chaotic with peaks in velocity dispersion exceeding $100$ km s$^{-1}${\it (right panels)}. In the next subsections we will discuss the \halpha morphology and kinematics, emission-line properties, and masses of the sample. For convenience IDs are shortened and referred to by their last three unique digits.
 
\subsection{Classification}
We initially classified the galaxies into three groups based solely on \halpha morphology: single emission, extended emission, and multiple emission. Galaxies classified as single-emission have a single resolved source of \halpha emission: {\it 16S, 14S, 18S, 46S, 52S}. Systems classified as extended emission show multiple unresolved or extended components: {\it 02S, 22S, 01R, 41S}.  Systems {\it 02R, 06R, 13R, 61S} are classified as multiple-emission galaxies, which have multiple, resolved regions of \halpha emission. The centres of the separate components are marked in Figure~\ref{fig.kinematics}.  These classifications are made for the purpose of discussing properties of galaxies with similar morphologies. The classifications are used to order Tables and Figures. Kinematic classifications are discussed in Section 4.

\subsection{Kinematic Quantities}
A flux-weighted velocity dispersion ($\sigma_{\mathrm{mean}}$) was calculated from the dispersion and flux across all spaxels.  We define velocity dispersion by \smean,~ which is representative of the internal motions of the gas \citep{2007ApJ...669..929L, Green:2010fk}. For comparison, a line-of-sight velocity dispersion ($\sigma_{\mathrm{net}}$) was calculated from the width of a Gaussian fit to the \halpha emission of the integrated spectrum. The measured values of \smean~are consistent with or slightly less than $\sigma_{\mathrm{net}}$,  as expected from velocity smearing. The values of \smean~ in this sample reach to $\sim130$ km s$^{-1}$, and are typically higher than in the samples of \cite{2009ApJ...697.2057L} and \cite{2009A&A...504..789E}. We note that the 0.15$''$ smoothing described in Section~2.3 produces a +10 km s$^{-1}$ median boost compared to \smean~calculated from the non-smoothed data which is less than the statistical error. We give the \smean~values from the smoothed data in Table~\ref{kinematics.table} for comparison to the literature. 

We do not make a correction based on beam smearing resultant from the PSF, which can be a problem for interpreting kinematic observations at high redshift (e.g. \citealt{2009ApJ...697.2057L, 2009ApJ...699..421W, 2009ApJ...706.1364F, 2010MNRAS.401.2113E}). Although all observations were taken in LGSAO mode, with a PSF of FWHM$\sim$850 pc, \cite{2009ApJ...697.2057L} and \cite{2009ApJ...699..421W} note that beam smearing can still have the effect of reducing the maximum rotation velocity and inflating the velocity dispersion for edge on disks in particular. However, as the inclination is not well constrained in this study it is unclear when and what correction would be appropriate. The same authors note that beam smearing does not explain the high velocity dispersions of objects which show no rotational velocity structure.

\begin{figure}
\includegraphics[scale=0.55]{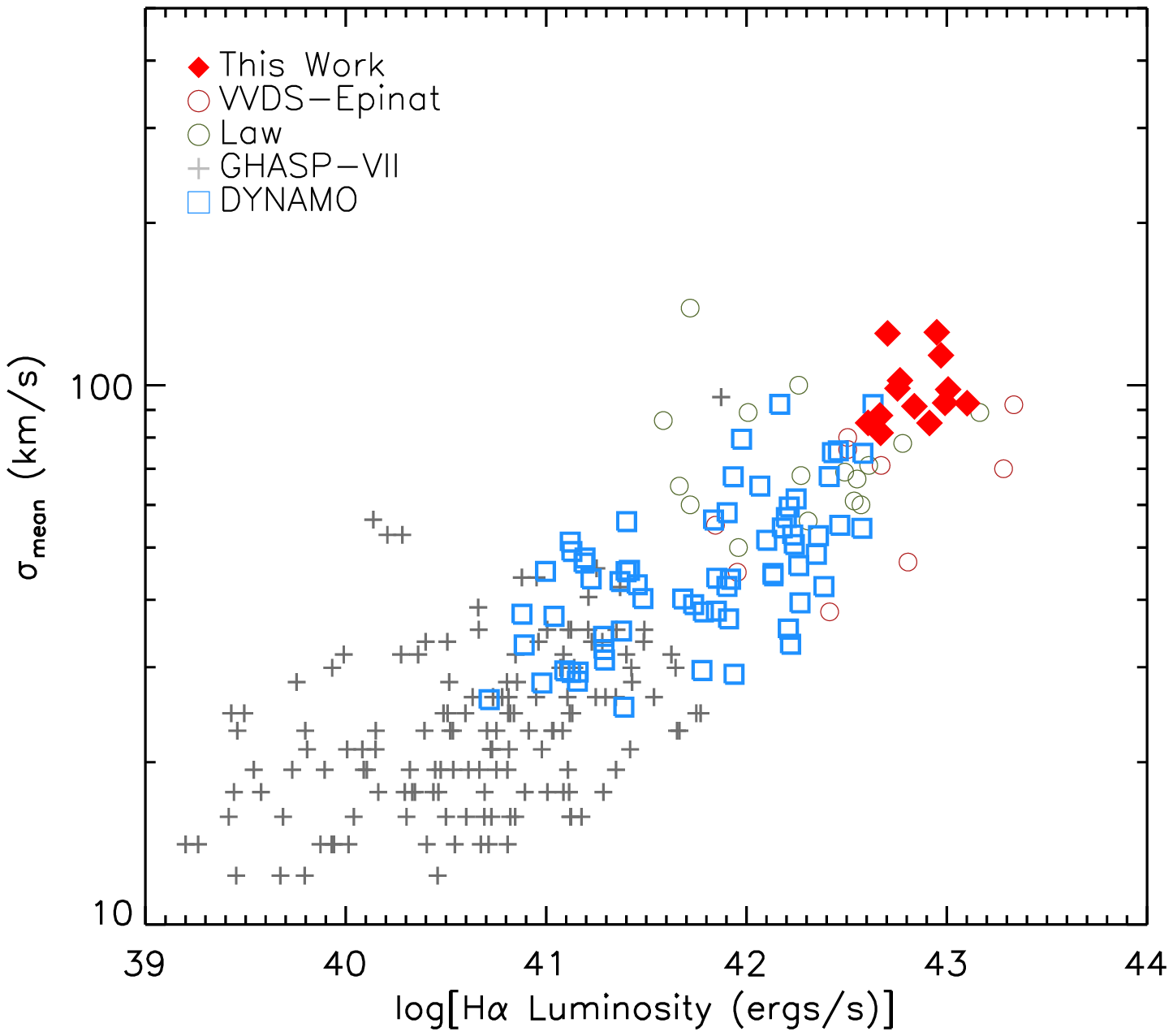}
\caption{
Flux weighted velocity dispersion as a function of $L_{H\alpha}$ from \citet{Green:2010fk} with galaxies from this sample as red diamonds. The WiggleZ points extend this relationship to higher \smean~and \halpha luminosity.\label{fig.andy}
}
\end{figure}

The observed dispersion and luminosity of the WiggleZ sample continues the relationship, spanning over $0<z<3$, observed by \cite{Green:2010fk} for star-forming galaxies, recreated in Figure~\ref{fig.andy}. The galaxies in the WiggleZ sample follow this relationship tightly with the highest dispersions and \halpha luminosities of the IFS samples introduced in Figure~\ref{fig.litcomp}. \cite{Green:2010fk} argue that the correlation seen implies star formation rate is the driving factor of the high velocity dispersion, not mass or gas fractions. Consequently, arguments that high velocity dispersions at high redshift are evidence for cold-flow accretion are weakened as the cold flows are no longer in place at the current epoch to produce the high dispersions observed in local galaxies. 

\begin{figure}
\includegraphics[scale=0.55]{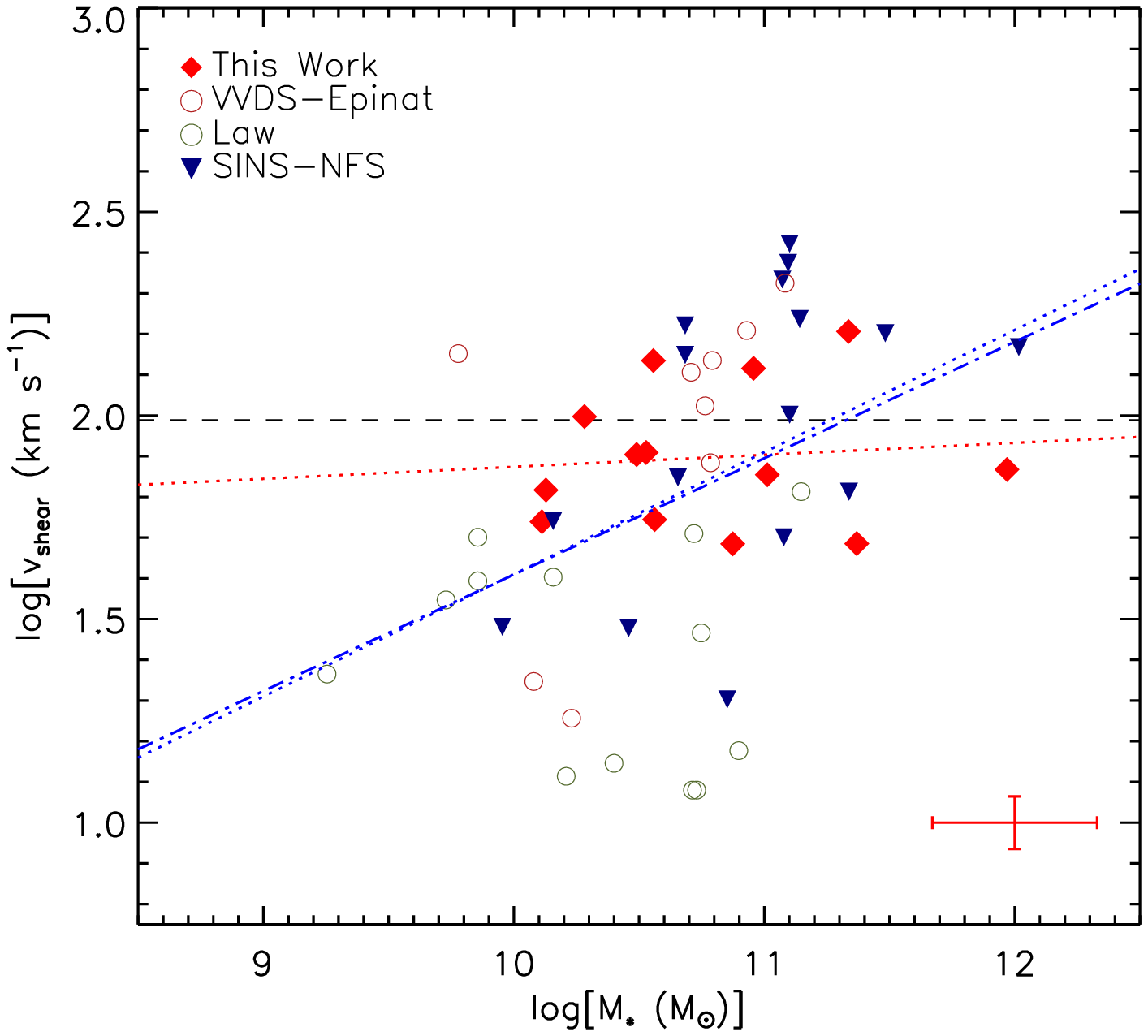}
\caption{
Logarithmic velocity shear as a function of logarithmic stellar mass for our 13 galaxies and galaxies from \citet{2009ApJ...697.2057L,2006Natur.442..786G,2008ApJ...687...59G,2006ApJ...645.1062F} and \citet{2009A&A...504..789E}. Data are not corrected for inclination. All stellar masses are converted to the Salpeter IMF for ease of comparison with the literature. A median error for the WiggleZ points is plotted in red in the lower right corner. The red dotted line is the best-fitting linear model to the 13 galaxies in our sample [$\log$(\vshear/km s$^{-1}$) $= 0.03\times\log$(M$_{*}$/\Msun)$+1.58$]. The blue dot-dash line is the best-fitting linear model to the data from all the samples [$\log$(\vshear/km s$^{-1}$) $=0.25\times\log$(M$_{*}$/\Msun)$-0.84$], the blue dotted line is the best-fitting linear model from \citet{2009ApJ...697.2057L} [$\log$(\vshear/km s$^{-1}$) $=0.29\times\log$(M$_{*}$/\Msun)$-1.25$] and the black dashed line is at the value of $\langle\log$(\smean)$\rangle$.}
\label{fig.vshear}
\end{figure}

A velocity shear was measured across each galaxy, $v_{\mathrm{shear}} = \frac{1}{2}(v_{\mathrm{max}}-v_{\mathrm{min}}),$ following \cite{2009ApJ...697.2057L}. In order to prevent biasing from outliers, $v_{\mathrm{max}}$ and $v_{\mathrm{min}}$ are calculated from the mean of the highest and lowest 5\% of the velocity map respectively \citep{2010ApJ...724.1373G}. The values of $v_{\mathrm{shear}}$ underestimate the true rotational velocity of the gas in the cases that the galaxy extends to greater radii than detected by OSIRIS.  The values of \smean, $\sigma_{\mathrm{net}}$, $v_{\mathrm{shear}}$, and $v_{\mathrm{shear}}$/\smean~are given in Table~\ref{kinematics.table}. We do not find a correlation between velocity shear and stellar mass in this sample of 13 galaxies, in disagreement with the $z\sim3$ sample of \cite{2009ApJ...697.2057L} and Lyman break analogs at $z\sim0.1$ \citep{2010ApJ...724.1373G}. However, in the context of other star-forming galaxies at $1<z<3$ a positive correlation is seen, as shown in Figure~\ref{fig.vshear}. The interpretation of the relationship is that stable rotation is more prevalent in galaxies that have assembled a relatively large stellar population. However, the addition of this sample does not affect the correlation originally published by \cite{2009ApJ...697.2057L}. It is unclear if the velocity shear from all samples are comparable given the different spatial scales probed by OSIRIS and SINFONI and the different signal-to-noise reached for each set of observations.

\subsection{Star Formation Rate Indicators}
Star formation rates were measured from WiggleZ \OII~and OSIRIS \halpha luminosities. Due to uncertainties arising in the flux calibration of optical fiber spectroscopy, the WiggleZ \OII~fluxes are only accurate to within 40\% due to systematic errors. OSIRIS \halpha flux measurements are comparably more accurate and are thus used as the primary measure of the nebular star formation.  Star formation rates (SFR) were calculated from \cite{1998ARA&A..36..189K},
\begin{eqnarray}
\mathrm{SFR}_{\mathrm{H}\alpha} [\mathrm{M}_{\odot}~\mathrm{yr}^{-1}] = \frac{1}{1.82}\frac{L_{\mathrm{H}\alpha}~[\mathrm{ergs~s}^{-1}]}{1.26\times10^{41}} \\ 
\mathrm{SFR}_{\mathrm{OII}} [\mathrm{M}_{\odot}~\mathrm{yr}^{-1}] = \frac{1}{1.82}\frac{L_{\mathrm{OII}}~[\mathrm{ergs~s}^{-1}]}{7.1\times10^{40}} 
\label{eq.av}
\end{eqnarray}
where the 1/1.82 converts to the BG03 IMF. 

Further measurement of star formation is derived from the SEDs generated from the stellar mass fitting described in Section 2.5 (SFR$_{\mathrm{SED}}$). A dust extinction, $Av_{\mathrm{SED}}$, is estimated from the SED fitting described in Section 2.6 and given in Table~\ref{stellar.table}. The mean error on $Av_{\mathrm{SED}}$ is 0.2. 

\begin{figure}
\includegraphics[scale=0.55]{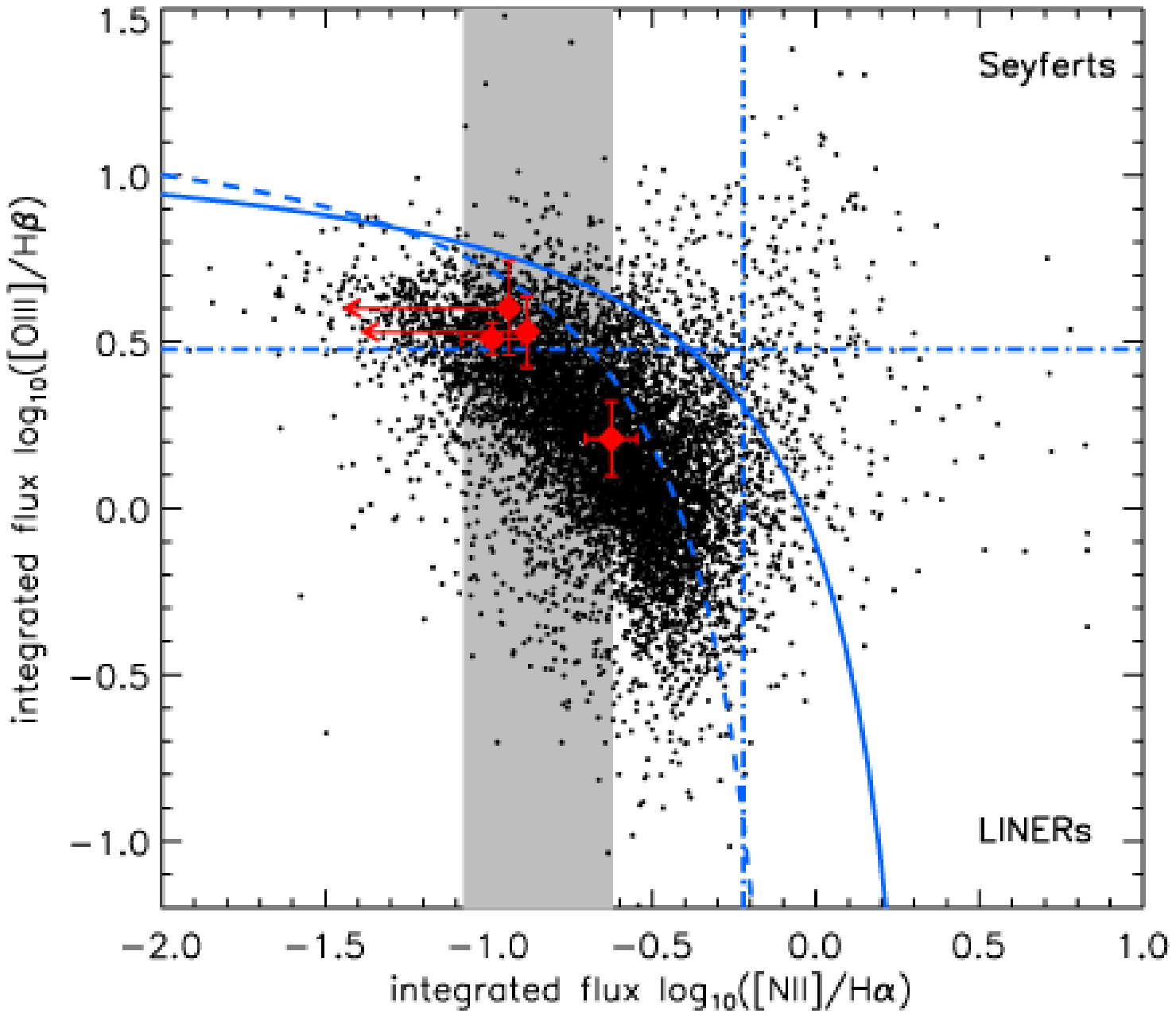}
\caption{
The emission line diagnostic diagram of \citet{1981PASP...93....5B}. The black points are WiggleZ galaxies and the red diamonds are {\it 52S}, {\it 22S}, {\it 61S} and {\it 18S} from this sample. The shaded gray area is the range containing the \NII/\halpha ratio for the remainder of the systems in this sample. The dashed curved line is the empirical line from \citet{2003MNRAS.346.1055K}, the solid curved line is the theoretical curve from \citet{2001ApJ...556..121K} and the vertical and horizontal lines are used to identify Seyferts and LINERS as labelled. The WiggleZ sub-sample presented here is primarily in the region of the diagram were star formation is the dominant mechanism driving line emission.
\label{fig.bpt}
}
\end{figure}

\subsection{Metallicity}
\NII~was detected in 8 of the 13 galaxies in our sample. The detections and upper limits are used to investigate the metallicity and level of contamination from hosting an AGN by using the \NII$\lambda$6584/\halpha ratio.
Metallicity was calculated from this ratio, as calibrated by \cite{2004MNRAS.348L..59P},
\begin{eqnarray}
12+\mathrm{log}(\frac{\mathrm{O}}{\mathrm{H}}) = 8.90 +0.57 \times \mathrm{log}(\frac{ \mathrm{\NII} }{ \mathrm{H}\alpha })
\label{eq.metallicity}
\end{eqnarray}
and given in Table~\ref{flux.table}. The metallicities of the sample are all sub-solar and are consistent with the results of IFS samples at $z\sim1-3$ of \cite{2008ApJ...687...59G,2008Natur.455..775S,2009ApJ...697.2057L,2009ApJ...706.1364F,2009ApJ...699..421W} and DEEP2 star-forming galaxies at $z\sim1.3$ \citep{2005ApJ...635.1006S}.  The typical errors in $12+\mathrm{log}(\frac{\mathrm{O}}{\mathrm{H}})$ are 0.2.

Metallicity measurements calculated from Equation~\ref{eq.metallicity} alone can be biased due to elevated levels of \NII~caused by AGN contamination. We look for this bias by constraining the position of the galaxies in this sample on the emission line diagnostic, or BPT, diagram \citep{1981PASP...93....5B}, shown in Figure~\ref{fig.bpt}. NIRSPEC observations of \OIII $\lambda$5007, 4959 and \hbeta of four galaxies in the sample further quantify AGN contamination using the \OIII $\lambda5007$/\hbeta ratio. 
Two of the \NII /\halpha measurements are upper limits (see column 5 of Table 2), and in all cases the \NII /\halpha ratios are significantly lower than the level expected for the given values of \OIII$\lambda5007$ / \hbeta if there was a significant component of AGN emission.
The remainder of galaxies all lie to the left of $\log$(\NII/\halphans) = $-0.6$, in the shaded gray region of Figure~\ref{fig.bpt}. Although AGN contamination can not be ruled out completely for these sources, their integrated nebular emission is clearly dominated by star formation. 

\begin{figure*}
\includegraphics[scale=0.68]{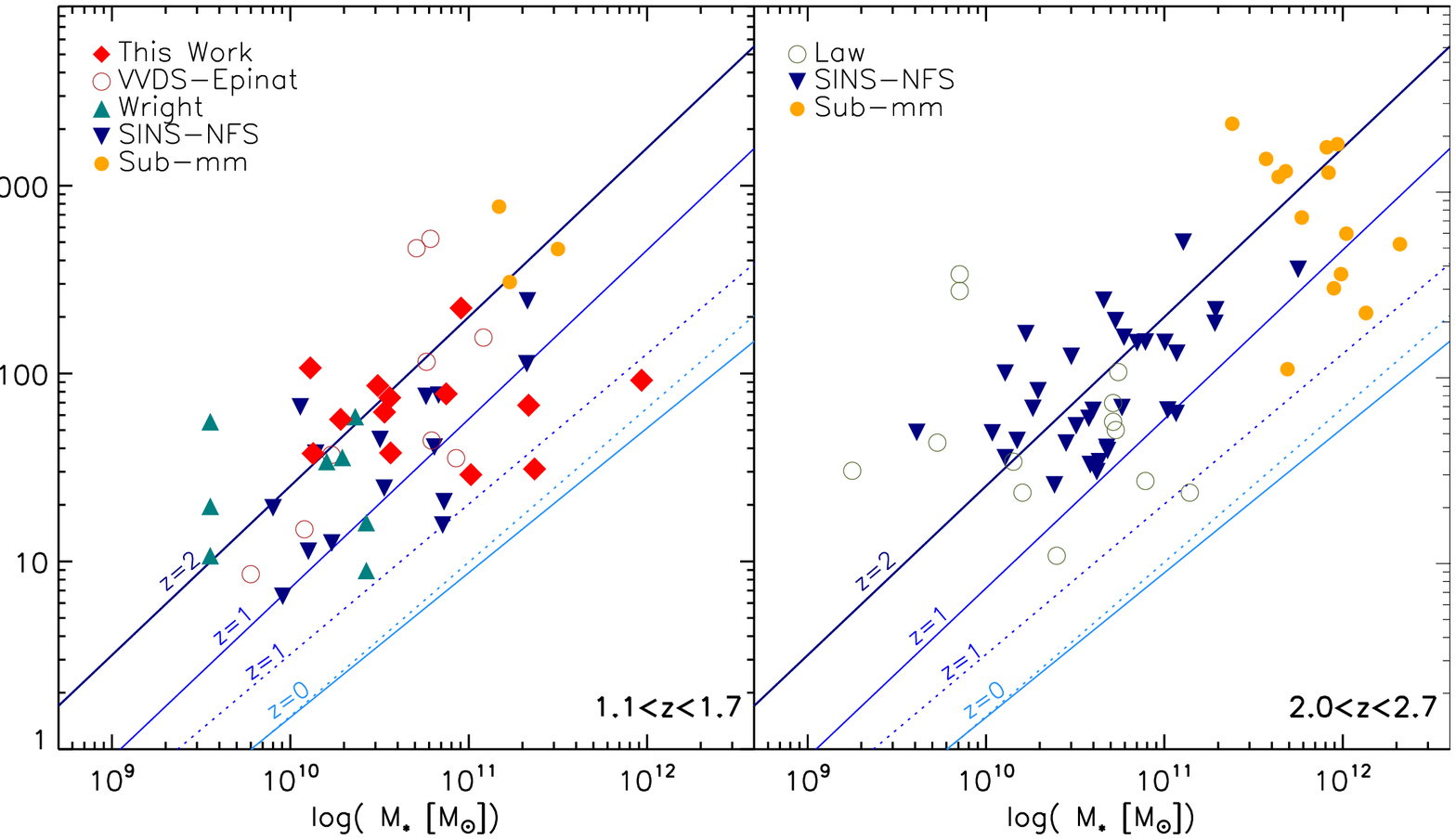}
\caption{
\halpha star formation rates of IFU samples at $z>1$ as a function of stellar mass. Stellar masses and SFRs for all samples are converted to the Salpeter IMF for ease of comparison to the literature. The symbols are as described in Figure~\ref{fig.litcomp}.   The solid lines are from the Great Observatories Origins Deep Survey (GOODS) given in \citet{2007A&A...468...33E} and \citet{2007ApJ...670..156D}. The dotted lines are from the Millennium simulation\citep{2005Natur.435..629S,2006MNRAS.365...11C} at redshifts $z=0$ (light blue) and $z=1$ (blue). An extinction correction is made from E(B$-$V)$_{\mathrm{SED}}$. 
\label{fig.ssfr}
}
\end{figure*}

\subsection{Mass Estimates} 
The sample stellar masses range from $9.8<\log(\mathrm{M}_{*} [\Msun])<11.7$, comparable to masses from other IFS samples covering the same redshift range ($1.2<z<1.7$; \citealt{2009A&A...504..789E,2009ApJ...699..421W,2009ApJ...706.1364F}) but with a higher $\langle$M$_{*}\rangle$ than high-redshift IFS samples ($z>2$; \citealt{2009ApJ...697.2057L,2008ApJ...687...59G}). No correlation is found between \halpha morphology and mass; the three most massive systems cover all three morphological classes, and the three lowest-mass galaxies cover two morphological classes.

With stellar mass estimates we are able to identify where these systems lie relative to the main sequence of star formation (SFR vs. M$_*$) as a function of redshift informing on the intensity of their star formation (e.g. \citealt{2007ApJ...660L..43N, 2010A&A...518L..25R}). 
If positioned above the main sequence, outside the limited range of SFRs at a given M$_*$, the star formation of the sample may be indicative of episodic variations rather than being associated with the larger population of continuous star-forming galaxies. 
Furthermore, the main sequence evolves with redshift in which higher SFRs were more typical in past epochs.  \cite{2010A&A...518L..25R} show that the mean specific SFR (SSFR$\equiv$SFR/M$_*$) of galaxies decreases by a factor of $\sim15$ from $z=2$ to $z=0$ for the most massive galaxies. 
The main sequence for IFS samples is shown, as SFR vs M$_*$, in Figure~\ref{fig.ssfr} in two redshift bins to account for this evolution. The SSFR is calculated, using SFR$_{\mathrm{H}\alpha}$, and presented in Table~\ref{stellar.table}. 

In the lower-redshift bin of Figure~\ref{fig.ssfr}, $1.2<z<1.7$, the star-forming galaxies of the SINS survey and \cite{2009ApJ...699..421W} form a sequence of star formation which scatter between the GOODS $z=1$ and $z=2$ empirical relations. The WiggleZ points in contrast do not show an increase in SFR with mass and therefore the galaxies in this sample below $\mathrm{M_{*}}=5\times10^{10}$~\Msun~scatter above the main sequence and the galaxies with masses $>5\times10^{10}$~\Msun~scatter below the main sequence. We note that the extinction correction applied here is derived from the SED fitting and future measurements of the dust content from far-infrared indicators will provide a more accurate correction affecting the location of this sample along the main sequence.

The high-redshift bin, $2.0<z<2.7$, shows a similar main sequence from the star-forming galaxies of the SINS Survey and \cite{2009ApJ...697.2057L} which scatters around the GOODS $z=2$ empirical relation. The mean SFR in the low-redshift bin ($\langle z\rangle\sim 1.52$) is a factor of 0.7 less than the the mean SFR in the high-redshift bin ($\langle z\rangle\sim 2.3$). Sub-millimeter galaxies (SMGs) are included in Figure~\ref{fig.ssfr} for comparison to the IFS samples. SMGs are considered as the `maximum starbursts'  \citep{2008ApJ...680..246T} with infrared determined SFRs $>1000$ \sfrunits. For consistency, only SMGs with published nebular emission are included in the comparison to IFS results \citep{2004ApJ...617...64S, 2006ApJ...651..713T}. SMG stellar masses are taken from \cite{2010A&A...514A..67M}. SFRs for all samples are calculated from measured nebular flux values with an extinction correction from $A_{\mathrm{V,SED}}$ and with a Salpeter IMF for consistency.  

We find that the SMGs in the low-redshift bin are well matched statistically to the lower mass data presented here and higher on average than the other IFS samples. In the high-redshift bin the SMGs are offset in mass from the IFS samples. Unfortunately, the low number of SMGs with measured line emission limits an in-depth comparison to this interesting galaxy population. However, initial comparisons to the available data indicate the WiggleZ sample presented here may be related to SMGs with more episodic star formation than LBGs at a given redshift and stellar mass.

\begin{figure*}
\setcounter{figure}{7}
\includegraphics[scale=0.85]{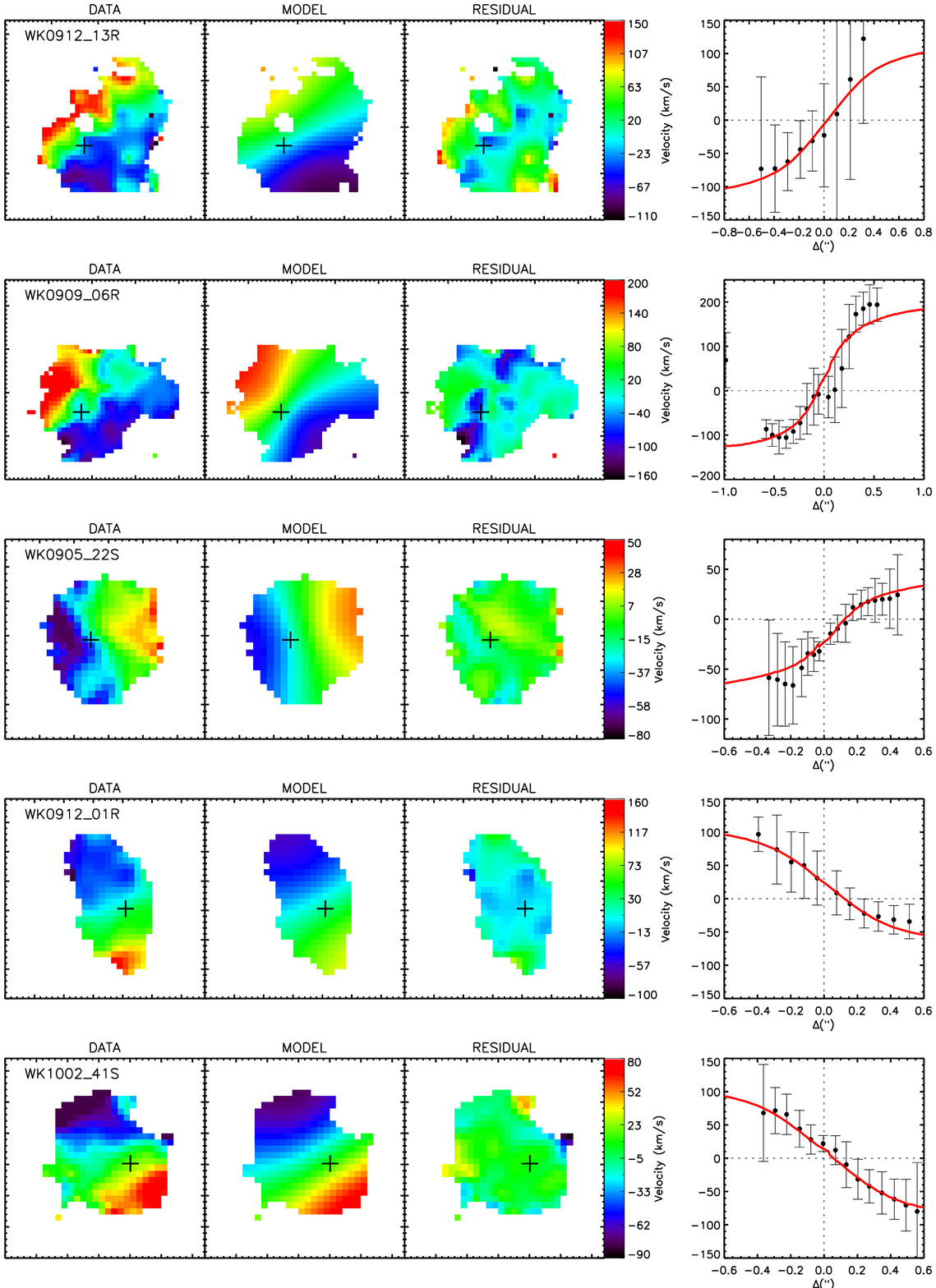}
\caption{
Disk fitting results for the 7/13 galaxies well-fit by disks in the sample. The first panel shows the smoothed velocity map, the second panel shows the smoothed modeled velocity map fit to the unsmoothed data, the third panel shows the residual from the difference of panel 1 and panel 2 and the final panel shows the binned rotation curve across the major axis with the model overlaid in red. 
\label{fig.models}
}
\end{figure*}

\begin{figure*}
\setcounter{figure}{7}
\includegraphics[scale=0.85, trim= 0mm 150mm 0mm 0mm, clip]{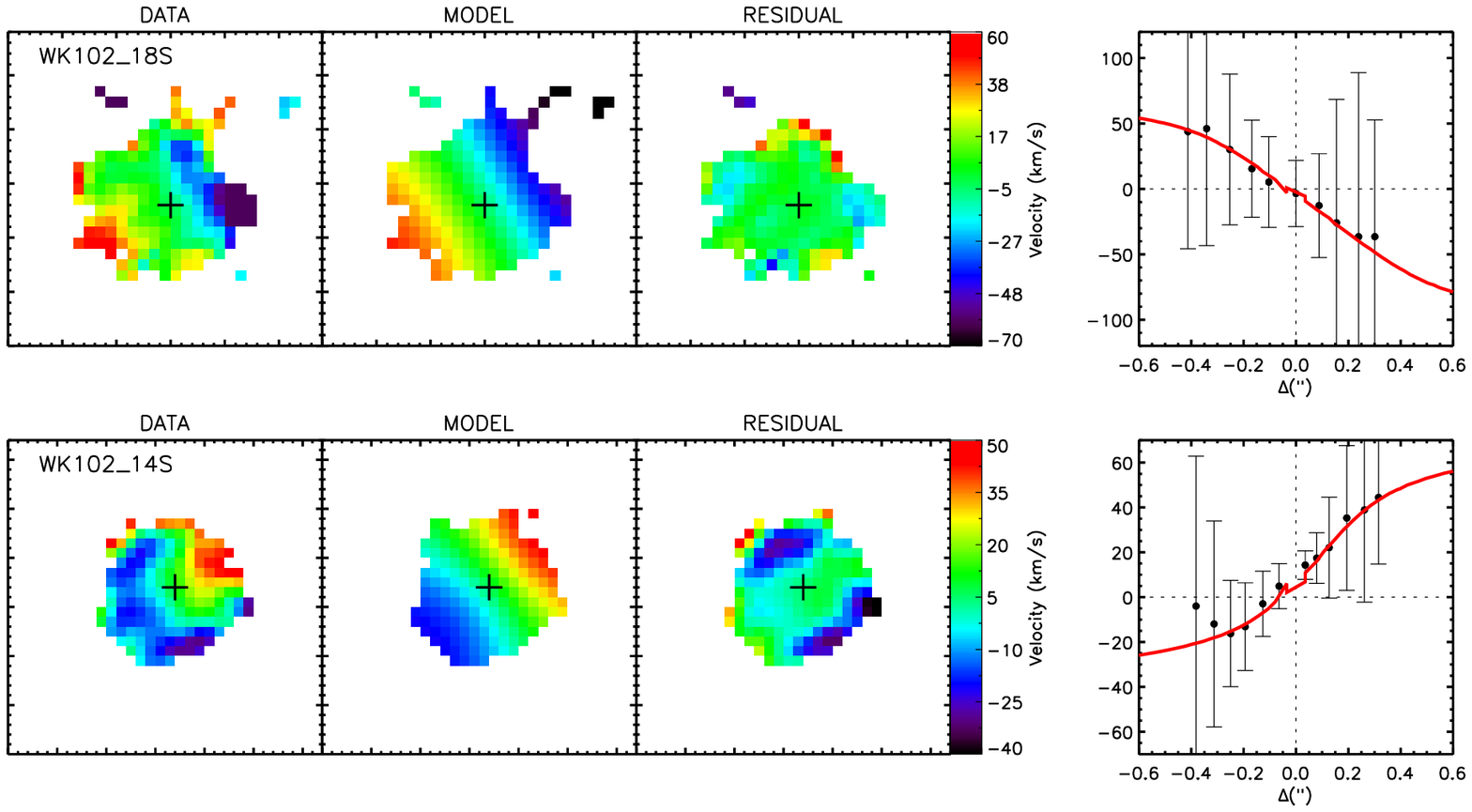}
\caption{
{\it cont.}
\label{fig.models}
}
\end{figure*}

\begin{table*}
\begin{minipage}{\textwidth}
\caption{Disk Model Parameter Results}
\begin{tabular*}{\textwidth}{@{\extracolsep{\fill}}lccccccc}
\hline
{ID} & 
{$\phi$} & 
{$V_\mathrm{p}\sin(i)$} &
{$r_\mathrm{p}$} & 
{$V_\mathrm{0}$} & 
{Residual} & 
{$\chi^2_\mathrm{R}$}  \\
{} & 
{} & 
{(km s$^{-1}$)} & 
{(kpc)} & 
{(km s$^{-1}$)} & 
{(km s$^{-1}$)} & 
{            }  \\
\hline
WK0912\_13R &       25$^{+331}_{-14}$  &       73$^{+64}_{-32}$ &        2$^{+20}_{-1}$ &       8$^{+14}_{-9}$ &      3.2 &       1.97 \\
WK0909\_06R &       53$\pm$2    &            187$\pm$7  &                1$\pm$1        &     43$^{+5}_{-6}$ &        4.3 &       5.12   \\
WK0905\_22S &       287$^{+10}_{-6}$  &       102$^{+80}_{-26}$ &        5$^{+3}_{-7}$  &      -7$^{+32}_{-22}$ &     3.5 &       2.27  \\
WK0912\_01R &       226$^{+21}_{-35}$  &       83$^{+49}_{-43}$ &       18$^{+6}_{-3}$  &     -14$^{+7}_{-8}$ &       2.6 &       3.44   \\
WK1002\_41S &       219$\pm$14    &          141$^{+126}_{-20}$ &       10$^{+6}_{-2}$  &      -7$^{+28}_{-26}$ &     1.5 &       0.99   \\
WK1002\_14S$^{a}$ &       320$\pm$36  &       42$^{+49}_{-30}$ &        17$^{+3}_{-17}$ &       15$^{+6}_{-5}$ &       1.1 &       1.71  \\
WK1002\_18S$^{a}$ &       140$^{+58}_{-65}$  &       24$^{+32}_{-21}$ &       16$^{+8}_{-3}$  &      28$\pm$7   &           2.1 &       2.58   \\
\hline
\end{tabular*}\\
{$^{a}$Kinematic center fixed to peak of \halpha flux distribution}.\\
\label{disk.table}
\end{minipage}
\end{table*}

\subsection{Disk Fitting}
The velocity maps of the galaxies were modeled to determine whether the data are consistent with disks.  Inclined disk models are generated varying the kinematic centre ($x_\mathrm{0}, y_\mathrm{0}$), the inclination ($i$, where $i$=0 is a face-on disk), the position angle ($\phi$), the systemic velocity ($V_\mathrm{0}$), the plateau velocity ($V_\mathrm{p}$), and the plateau radius ($r_\mathrm{p}$) using the prescription of \cite{1989A&A...223...47B},

\begin{eqnarray}
V(x,y) = V_\mathrm{0} + V_\mathrm{c}(r)\sin(i)\cos(\theta),
\label{eq.v1}
\end{eqnarray}
where $V_\mathrm{c}$ is the circular velocity at a given radius from ($x_\mathrm{0}, y_\mathrm{0}$) and $\theta$ is the azimuthal angle in the plane of the galaxy, defined as:
\begin{eqnarray}
&&\cos(\theta) = \frac{-(x-x_0)\sin(\phi)+(y-y_0)\cos(\phi)}{r},\nonumber\\
&&\sin(\theta) = \frac{-(x-x_0)\cos(\phi)+(y-y_0)\sin(\phi)}{r\cos(i)}\label{eq:xdef}.
\label{eq.v2}
\end{eqnarray}
The circular velocity is defined as: 
\begin{eqnarray}
V_\mathrm{c}(r) = \begin{cases}
& V_\mathrm{p}\frac{r}{r_\mathrm{p}}, \hspace{0.7cm} r\leq r_\mathrm{p}\\ 
& V_\mathrm{p}, \hspace{1cm} r> r_\mathrm{p} 
\end{cases}
\label{eq.v3}
\end{eqnarray}
for a flat rotation curve \citep{2007ApJ...658...78W}. The `flat' disk model has been shown statistically to reproduce fitting parameters more successfully than the exponential disc, isothermal sphere, and arctangent models in artificially redshifted IFS data \citep{2010MNRAS.401.2113E}. The models are convolved with the observed PSF, modeled from a tip-tilt star observation, and are fit to the data prior to the 0.15$''$ smoothing to reduce the effects of beam smearing. The best-fitting models were determined by minimizing the reduced chi-squared value ($\chi^{2}_\mathrm{R}$) using Monte Carlo Markov Chain methods.  The parameter results of the best-fitting models, $\chi^{2}_\mathrm{R}$ values, and the average residual velocity are given in Table~\ref{disk.table}. Example models are shown in Figure~\ref{fig.models}. The residual velocity, $v_{\mathrm{res}}$, is calculated from the pixel average of the absolute value of the difference of the data and model. 

The kinematic center is not well constrained in single emission galaxies, particularly in the single emission galaxies that show no shear or have fewer spatial pixels. In these cases, specified in Table~\ref{kinematics.table}, the kinematic center is fixed to the peak of \halpha emission. Although the peak of \halpha emission is not necessarily the dynamical center of the galaxy, \cite{2009ApJ...699..421W} find that it is a reasonable approximation.

We find that 7/13 of our galaxies ({\it06R, 01R, 13R, 41S, 22S, 14S, 18S}) are well fit by a disk model. To determine which objects are well fit by this simple disk model we consider the velocity residual, reduced chi-square value and a qualitative measure of coherent velocity shear. We note that $\chi^{2}_\mathrm{R}$ is biased as the spaxels are correlated due to the intrinsic PSF and is not alone a reliable indicator. Furthermore, small fluctuations in the velocity map resulting from galactic winds and/or interactions between individual star-forming regions will inflate the residual and reduced chi-squared values as they are not accounted for in the model. 

A known approximate degeneracy exists between $V_\mathrm{p}$ and $i$ in the form $V_\mathrm{p}\sin(i)$ in Equation~\ref{eq.v1} (e.g. \citealt{1989A&A...223...47B, 2009ApJ...697.2057L,2009A&A...504..789E}). In previous low signal-to-noise IFS work the inclination is set to a constant of $i=45$ \citep{2009ApJ...699..421W} however, \cite{2009ApJ...697.2057L} derive $i=57.3$ as the correct average inclination between our line of sight to a given disk and the vector normal to the disk. The most robust method may be to fix the inclination to values derived from deep broadband imaging \citep{2009A&A...504..789E}, however we consider $V_\mathrm{p}\sin(i)$ as one parameter in this study as we lack sufficiently deep imaging to determine it from photometry. 

We find another degeneracy present in {\it 41S}, {\it 14S} and {\it 13R} between the plateau radius and plateau velocity. This degeneracy arises from the linear nature of the velocity gradients observed and is likely a consequence of not probing to faint enough surface-brightness emission to detect the kinematic turnover. A characteristic peak of velocity dispersion, a result of rapid unresolved change in velocity, would not be expected in the dispersion maps. Similarly, face-on and low mass galaxies which have a low velocity gradient show a very faint or no central peak of velocity dispersion \citep{2010MNRAS.401.2113E}. As a result we do not use the central peak of velocity dispersion as an indicator for disk classification.

As found in previous IFS studies of high-redshift galaxies the v$_{\mathrm{shear}}$/$\sigma_{\mathrm{mean}}$ (Table~\ref{kinematics.table}) values of this sample are approximately unity. The $v/\sigma$ measurement is often used to make a split between rotationally supported systems ($v/\sigma>1$) and systems dominated by dispersion ($v/\sigma<1$). In this sample {\it 02R, 06R, 01R, 13R} have $v/\sigma>1$ while {\it 02S, 22S, 16S, 46S, 52S} have $v/\sigma<1$ and {\it 41S, 18S, 61S} have $v/\sigma\approx1$ within the errors. This correlates with the results of disk modeling in which {\it 06R, 01R, 13R} and {\it 41S} are well fit by disks and {\it 02S, 16S, 46S, 52S} are not well fit by disks. However, this picture fails for {\it 22S, 14S} which are well fit by disks but have $v/\sigma<1$ and {\it02R} which likely has an elevated $v/\sigma$ as a result of an outflow rather than from the signature of rotational support (See Appendix). It is likely that the discrepancy between these two methods is a result of the detection limit of the observations. The v$_{shear}$ measurement probes only the inner star-forming regions of the galaxies and is not representative of the overall rotational velocity of the galaxy, thereby reducing the value of $v/\sigma$. Other factors are known to reduce the observed values of $v/\sigma$ such as the foreshortening due to the inclination by a factor of $\sin i$ \citep{2009ApJ...697.2057L}. Because the inclination values in this study are poorly constrained this correction is not applied. As a result, more weight is given to the results of disk modeling rather than $v/\sigma$ in classifying galaxies as being rotationally supported.

\subsection{Surface Brightness Profiles}
The effective radius was calculated by fitting convolved 1-D S\'ersic models to the \halpha surface brightness profiles of single-emission galaxies. In order to fit model S\'ersic profiles, the central spaxel of the galaxy or star-forming complex was first identified from the peak of the flux maps. The distance to the remaining spaxels was calculated from the central point assuming a circular profile, reducing the number of free parameters by setting the aspect ratio to unity and position angle to zero. Individual measurements are averaged in radial bins.  The models were convolved with the PSF. The best-fitting S\'ersic profile model was determined by minimizing the chi-squared value ($\chi^2$) with three free parameters, $n$, $r_{0}$, and  {\it $\Sigma_0$}, where $n$ is the S\'ersic index, $r_0$ is the effective radius and {\it $\Sigma_0$} is the surface brightness at the effective radius. We assume this simple model as the inclination and position angle determined in Section 3.6 are not well constrained.

\begin{figure}
\includegraphics[scale=0.55, viewport=10 5 600 390,clip]{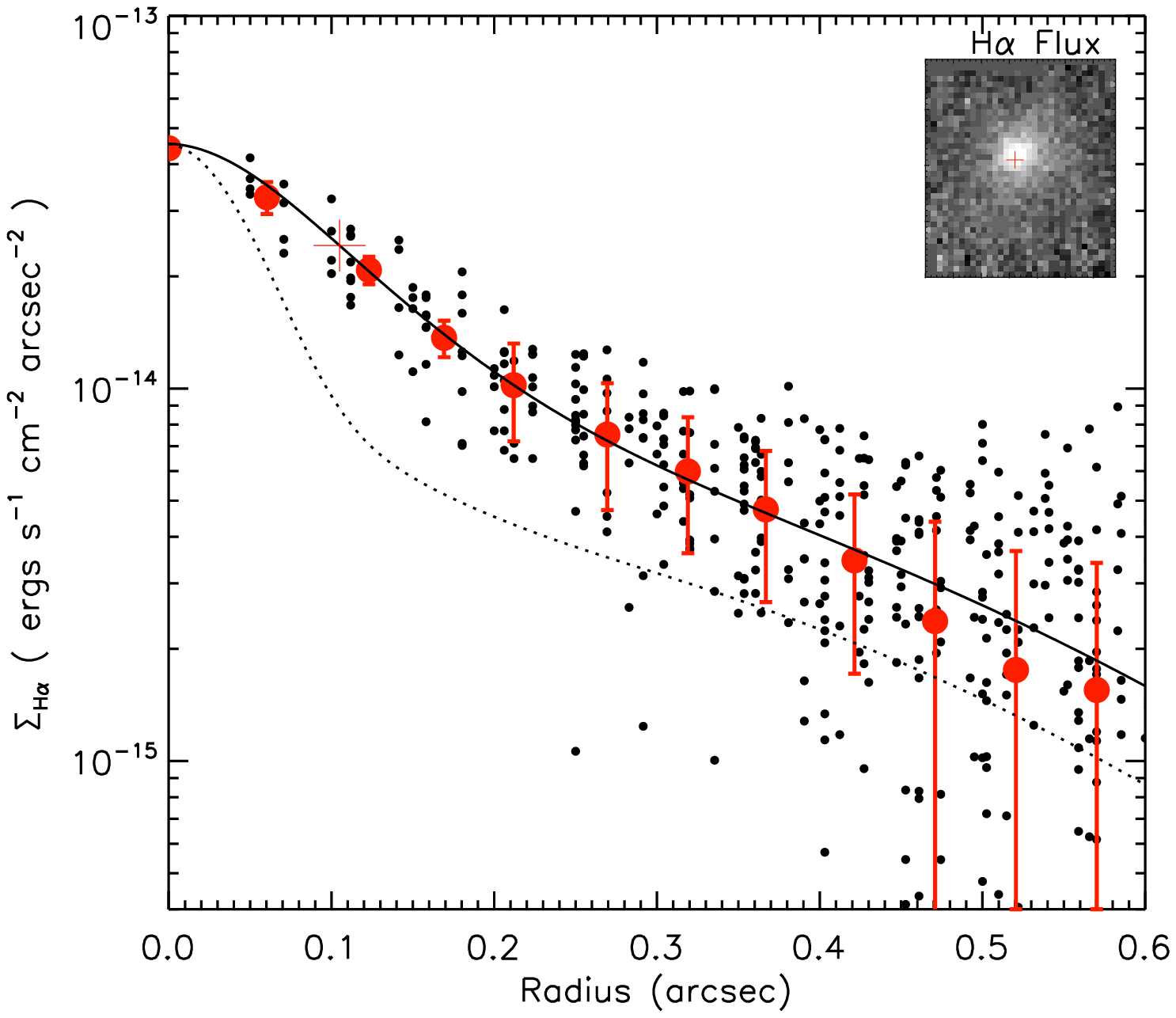}
\caption{
Example of surface brightness modeling results for {\it 14S}. The black points represent the flux in each spaxel from the peak of emission marked with the red plus sign in the \halpha image of the galaxy in the top right corner. The red circles represent binned points, the black solid line is the best-fitting S\'ersic model, and the black dotted line is a fit to the flux profile of a tip-tilt star observed with OSIRIS.
\label{fig.sb}
}
\end{figure}

The fitted results are shown in Table~\ref{sb.table} with $r_{\mathrm{eff}}$ ranging from $0.4$ kpc $<r_{\mathrm{eff}}<1.9$ kpc and $n$ ranging from $0.5<n<2.0$. An example model fit to the data is shown in Figure~\ref{fig.sb}. It is not completely understood what the flux profile of \halpha emission reveals about the parent galaxy as S\'ersic profiles are more commonly fit to broadband emission representative of continuum emission from continuous star formation \citep{1968adga.book.....S}. However, the SINS survey obtained deep rest-frame optical images of six irregular galaxies in their sample and found surprisingly good morphological agreement between \halpha and broadband images \citep{2010arXiv1011.1507F}. They find the S\'ersic parameters of \halpha emission and $H$-band emission (rest-frame optical) are consistent for the six galaxies with a difference on average of 0.08 for the S\'ersic index and agreement within 3\% for $r_{\mathrm{eff}}$.  The S\'ersic indices are comparable to our \halpha results with $n<1$ in all cases however they find greater effective radii with $r_{\mathrm{eff}}>2$ kpc in all cases, most likely due to the coarser resolution from their non-AO corrected observations.

The fitting method described above is applied to available $r$-band images from RCS2 only, as galaxies are unresolved in SDSS images.  The RCS2 images, taken in average seeing of 0.7$''$, are sufficient to fit S\'ersic profiles to 3 of 13 systems of the sample. We obtain results in the range $0.2<n<1.0$.  A direct comparison to the \halpha S\'ersic parameters however is not possible as the only RCS2 data available is of extended or multiple emission galaxies. Deeper imaging of the whole sample is needed to reveal whether the underlying stellar population probed by the rest-frame optical for these galaxies follows the morphology of the \halpha emission.
 
\begin{table}
\begin{minipage}{\textwidth}
\caption{Surface Brightness Fitting}
\begin{tabular*}{0.47\textwidth}{@{\extracolsep{\fill}}lcc}
\hline
{ID	} & 
{n } & 
{$r_{\mathrm{eff}}$(kpc)} \\
\hline
\vspace{0.15mm} \\
\multicolumn{3}{c}{\halpha Images}\\
\hline
\vspace{0.25mm} \\
WK0909\_02R & $1.0^{+0.1}_{-1.6}$    &  0.6$^{+0.2}_{-1.6}$  \\
WK1002\_18S & $0.8\pm0.1$  &  $1.9\pm0.4$   \\
WK1002\_14S & $0.9^{+0.03}_{-0.1}$ &  0.9$^{+0.1}_{-0.2}$   \\
WK1003\_46S & $1.0\pm0.1$   &   0.7$^{+0.1}_{-0.2}$   \\
WK1003\_52S & $2.0^{+0.6}_{-0.1}$  &  $0.4\pm0.1$   \\
WK0912\_16S & $0.5^{+0.1}_{-0.3}$  &  1.5$^{+0.2}_{-0.6}$   \\
 \hline
 \vspace{0.15mm} \\
 \multicolumn{3}{c}{RCS2 rest-frame UV Images}\\
 \hline
WK0912\_13R  &  $0.6^{+0.1}_{-1.8}$ & $4.6^{+2.9}_{-2.5}$ \\ 
WK0909\_02R  &  $1.0^{+0.2}_{-0.6}$ & $6.1\pm2.0$  \\
WK0909\_06R  &  $0.4\pm0.3$ & $3.4^{+2.6}_{-1.1}$  \\
\hline
\end{tabular*}
\label{sb.table}
\end{minipage}
\end{table}

\subsubsection{Radii}
We note that radius is a particularly difficult property to measure robustly and consistently with a variety of methods used in the literature. We calculate the kinematic plateau radius, $r_p$, from disk modeling and the effective radius, $r_{\mathrm{eff}}$, from surface brightness profile fitting.
The $r_{\mathrm{eff}}$ values presented are consistent with the sample of \cite{2009ApJ...697.2057L}, $0.6$ kpc $\leq r\leq1.6$ kpc, but are smaller than most other samples (e.g. \citealt{2009ApJ...699..421W}: $2.1$ kpc $\leq r\leq3.9$ kpc, \citealt{2009ApJ...706.1364F}: $1.2$ kpc $\leq r\leq7.5$ kpc, \citealt{2009A&A...504..789E}: $1.2$ kpc $\leq r\leq5.5$ kpc). The half light radius, $r_{1/2}$, is also estimated directly from the data by determining the size of the circular aperture needed to enclose half the \halpha light from the centre of emission, and is given in Table~\ref{kinematics.table}. In some cases $r_{1/2}$ is the best measure of radius as neither $r_{\mathrm{eff}}$ or $r_p$ are able to be measured due to irregular morphologies and kinematics. 
The $r_{1/2}$ values are greater on average than the $r_{\mathrm{eff}}$, likely due to noise, and are comparable to the samples of \cite{2009ApJ...699..421W} and \cite{2009A&A...504..789E}.

\section{Discussion}
The sample presented here was selected from the WiggleZ survey for strong UV and nebular luminosities. The luminosity, mass, SSFR, and $v_{\mathrm{shear}}$ are more comparable to the $z>2$ sub-set of the SINS survey than the samples at the same redshift, but more comparable in size to the $z\sim2$ sample from \cite{2009ApJ...697.2057L}. Like other IFS surveys we find a range of kinematics and morphologies with high velocity dispersions on average ($\langle\sigma_{\mathrm{mean}}\rangle=98$ km s$^{-1}$), yet the uniqueness of this sample is derived from the range of \halpha morphologies with resolved $kpc$-size star-forming regions and ultra-compact \halpha emission of the single emission galaxies. 

However, IFS samples are all selected differently and may not be directly comparable. The SINS survey observes galaxies found from a variety of selection techniques including optical color selected `BX/BM' galaxies \citep{2004ApJ...604..534S}, infrared color selected `BzK' galaxies \citep{2004ApJ...617..746D}, Lyman break selected galaxies (LBGs), and bright sub-millimeter galaxies (SMGs). 
\cite{2009A&A...504..789E} select their galaxies from the magnitude complete VVDS survey, while \cite{2009ApJ...699..421W} observe BM galaxies, and \cite{2009ApJ...697.2057L} observed BX galaxies.
While the WiggleZ sample was not selected by any of these methods, the SEDs of the sample imply that they could be classified as BM galaxies, BzKs and LBGs. Because IFS surveys are observationally expensive we are working with low number statistics and we cannot follow up mass or luminosity complete samples at high redshift. 

The initial classification of this sample was based on \halpha morphology into three groups; single emission, extended emission, and multiple emission.  A reclassification is now considered based upon the properties discussed in Section~3 in the context of disk formation at high redshift and a merger sequence.
However, it is a challenge to distinguish between a galaxy merger and the merger of massive sub-clumps embedded within a rotating disk and similarly to distinguish between a merger remnant and a pseudo-bulge formed from the collapse of massive sub-clumps. 
Below we explore models of both disk formation and merger events at high redshift.  We use the kinematic properties, masses, sizes, etc. to differentiate between the dominate formation mechanism for the WiggleZ galaxies sample presented here.

\subsection{The Merger Hypothesis}
With separations of 3--10 kpc the \halpha complexes may be sites of multiple regions of star formation merging from different parent galaxies. Kinematic quantities need to be utilized to find support for this hypothesis as deep imaging of diffuse continuum emission is unavailable to reveal tidal features characteristic of mergers. We analyze here a scenario in which our sample form a merger sequence with pre$-$, ongoing, and post$-$mergers. 

In this scenario the multiple emission galaxies are the initial stages of mergers when star formation in separate galaxies can still be spatially differentiated. The extended emission galaxies are the next stage in the sequence representing ongoing merging activity (or pre-mergers with component confusion due to projection effects). The final stage in the proposed merger sequence is the single emission galaxies representing merger remnants with high dispersions and star formation rates.

Velocity dispersion maps provide some evidence indicating that we are seeing separate kinematic components merging. The observed peaks in dispersion may correspond to centres of individual galaxies or be a result of shock fronts between merging systems. Systems {\it41S} and {\it61S} show multiple peaks in velocity dispersion correlated with the brightest regions of emission, indicative that the system is an early-stage merger where each galaxy has a peak in velocity dispersion corresponding to an unresolved component. Systems {\it13R} and {\it06R} show multiple peaks in velocity dispersion correlated with the gas between the brightest regions of emission, indicative of the turbulence occurring from the interaction of the gas from the merging galaxies in this scenario or the effect of fitting substructure along the line of sight with a single Gaussian. 

However, the majority of IFS results classified as mergers show velocity maps where the velocity does not vary smoothly across the source but rather show a quick turnover within a resolution element or less \citep{2008A&A...479...67N, 2009ApJ...699..421W, 2009ApJ...697.2057L}. An ideal example, presented in \cite{2008A&A...479...67N}, is a merging system observed as two unresolved ``clumps'' of \OIII~emission with a step in the velocity map between both components. This step function, observable with AO resolution, is not present in the kinematics presented here.

The single emission galaxies in this scenario could be merger remnants. This comparison is particularly difficult as merger remnants produced in simulations show a wide range of kinematic and morphological properties with observables sensitive to the initial mass ratio of the merger \citep{2003ApJ...597..893N}. However, the majority of remnants are slow rotators (with v/$\sigma<1$), and have high velocity dispersions and minor axis rotation \citep{2003ApJ...597..893N,2006ApJ...650..791C,2010MNRAS.406.2405B}. These properties could be used to describe {\it 22S, 16S, 14S, 18S} and {\it 46S} (although major and minor axes cannot be distinguished). These galaxies show some indication of rotation but with very low amplitudes (\vshear~$\sim$40 km s$^{-1}$). But it is difficult to distinguish merger remnants and disks. Some simulations even find that a low percentage of remnants also have disk-like  isophotes \citep{2003ApJ...597..893N}.

A fraction of the sample, particularly the lower-mass galaxies, are found above the `main sequence' of star formation (Figure~\ref{fig.ssfr}), which indicates they may be undergoing a burst of star formation typical of a gas-poor or minor merger event \citep{2007ApJ...660L..43N}. 

Unfortunately, surface brightness detection limits of OSIRIS may introduce a selection effect of detecting only the brightest objects in merging systems. This effect can be seen in Figure~A1b of the Appendix for object {\it 02R} where an outflow or companion is seen in the RCS2 image but not detected in the \halpha OSIRIS image. Classification errors may arise as a result of this effect. Deep broadband imaging of all sources will help to reveal merging companions.

\subsection{The Clumpy Disk Hypothesis}

We investigate the scenario where the data presented here represent different stages of the clumpy disk models of the \cite{Elmegreen:2008fk} simulations, hereafter EBE08. In this context galaxies with multiple regions of emission represent the early phases of disk formation when gravitational instabilities result in multiple star-forming complexes forming under Jeans collapse embedded within a disk galaxy. As turbulent speeds decrease relative to the rotational speed of the disk, the clumps migrate towards the centre on a timescale of 1 Gyr, represented here by the galaxies with extended emission. Finally, the clumps coalesce and form a slowly rotating pseudo bulge \citep{1999ApJ...514...77N,2007ApJ...670..237B,Elmegreen:2008fk,2009arXiv0901.2458D, Krumholz:2010fk}, represented here by galaxies with a single spherical emission profile.

The morphologies of multiple emission systems, {\it06R}, {\it13R}, and {\it61S} (Figure~\ref{fig.kinematics}) are well matched to the early stages of clumpy disk formation ($t=175-450$ Myr). Systems {\it06R} and {\it13R} are well fit by disk models.  
The clump velocities follow the global velocity structure of the galaxy as seen in Figure~\ref{fig.models}, where the major axis of rotation cuts across multiple clumps revealing no corresponding deviation in the rotation curves \citep{2004ApJ...611...20I}.

The morphologies of  {\it02S}, {\it22S}, {\it41S} and {\it01R} are either consistent with the system being observed at the clump coalescence phase ($t=600-800$ Myr) or the result of projection effects. Systems {\it02S} and {\it22S}, each show two close components which can be distinguished when fitting surface brightness profiles, while {\it41S} and {\it01R} have single bright extended regions where multiple components cannot be differentiated visually or from surface brightness profiles.  {\it22S, 41S} and {\it01R} are well fit by disk models whereas {\it02S} has a flat velocity map with a peak in dispersion corresponding to the brightest peak in \halpha emission. The fainter peak of emission may be a signature of a galactic outflow resulting in confusion in classification. 

Star-forming regions within a disk should have similar or consistent metallicities as they are evolving within the same conditions as predicted by theoretical models for disks at high redshift \citep{2004A&A...413..547I}.  In a major or minor merger no similar correlation is expected in the metallicities of each galaxy. By integrating the light of each clump, the clump metallicities are calculated from Equation~\ref{eq.metallicity}. Metallicities of clumps in {\it06R} and {\it61S} are equal within the errors, as predicted for the early stages of disk fragmentation. At clump coalescence metallicities of individual clumps would be consistent as seen in {\it02S} and {\it22S}. Although this supports current theories, $12+\log$(O/H) varies little across the sample ($\Delta[12+\log$(O/H)]$\sim0.26$). 

Compact single emission galaxies {\it14S}, {\it16S}, {\it18S}, {\it46S}, and {\it52S} have morphologies well matched to the later stage of simulated formation when the clumps have collapsed to form a bulge ($t=1000$ Myr). 
The dispersions of {\it14S}, {\it46S}, and {\it52S} are the highest in the sample and are consistent with EBE08 bulges, predicted to form by clump coalescence and become slow rotators with high turbulence of $\sigma_{\mathrm{los}}\sim130$ km s$^{-1}$ in their inner regions. Systems {\it14S, 16S, 18S} all show velocity gradients with amplitudes less than 70 km s$^{-1}$. The bulge density profiles in the EBE08 models are well fit by S\'ersic profiles of $n\sim4$ in the inner 500 pc and $n>2$ at greater radii. While we do not resolve the inner 500 pc we do resolve to $650-750$ pc and find $n<2$ in all cases for the \halpha emission. 

Velocity dispersion maps in this sample do not show characteristic central peaks in velocity dispersion observed in IFS studies in both local \citep{2002MNRAS.329..513D} and high redshift \citep{2006Natur.442..786G, 2009ApJ...706.1364F} disk galaxies and therefore cannot be used as an indicator for disk classification.  As discussed in Section 3.6, this peak in velocity dispersion can be attributed to beam-smearing across the inner regions of the velocity turnover of the galaxy. However, because we are observing only the central region of rotation and have a small PSF as a result of using LGSAO, a dispersion peak is not expected. 

Furthermore, if we model the velocity dispersion using the equation for a thick or compact disk \citep{2008gady.book.....B};
\begin{eqnarray}
\sigma \sim \frac{h_\mathrm{z}V_\mathrm{c}(r)}{r}
\label{eq.sigma}
\end{eqnarray}
where $h_\mathrm{z}$ is the disk scale height, the expected model would give a constant, featureless dispersion map within the plateau radius ($r\le r_\mathrm{p}$). This would be consistent with galaxies {\it41S, 14S}, and {\it13R} where a plateau velocity is never reached. In this case the observed variations in velocity dispersion are more likely a result of random perturbations within the centre of the disk or local perturbations due to clumps.

The simulations of EBE08 and others that model and evolve these galaxies have mainly been developed with sticky-particle codes \citep{Elmegreen:2008fk,1999ApJ...514...77N} which only model the cold gas associated with the cold cloud medium.  \cite{2004A&A...413..547I,2004ApJ...611...20I} do include a hot phase to their simulation allowing the galaxy models to track stellar ages and metallicities providing observables that can be tested. Furthermore, the initial conditions of such models require the disks to have very high velocity dispersions to mimic observational results at these redshifts making it difficult to distinguish inputs and outputs of the models. More realistic models are needed for a more rigorous comparison. Regardless of these limitations the data is well-matched to the available models.

\section{Conclusion}
We favor the clumpy disk model over the merger sequence model. We find 7/13 of the sample is well modeled by a disk and we find that clumps found in galaxies with multiple regions of emission do not strongly affect the kinematic fields. This provides observational evidence for the models of \cite{Elmegreen:2008fk} which provide an explanation for the irregular morphologies seen at high redshift. 

However, whilst natural to attempt to identify kinematic observations as disks or mergers, as been attempted here, this simplified picture of galaxies becomes increasingly fallible at high-redshift. More than half of the galaxy kinematics in this sample are well-described by disk models, however it is important to be cautious of definitive kinematic classifications. For example, \cite{2010MNRAS.401.2113E} provide evidence that up to 30\% of disk galaxies may be misclassified as perturbed rotators or complex kinematics with current classification schemes. 

It is likely that we are seeing a variety of mechanisms at work which cannot be broken down into simply disks and mergers. Even if disk instability is a dominant mechanism in star-formation histories at high redshift, as we find here, the disks are still likely to undergo mergers in their lifetime further complicating kinematic interpretations.  \cite{2010MNRAS.401.1657L}, for example, observe a galaxy pair with separation of 12 kpc with one galaxy in the pair classified as a disk with velocity gradient of $\sim$91 km s$^{-1}$ and the other galaxy classified is compact and likely dominated by AGN. It is unclear how to classify this system, which has multiple kinematic components.
Furthermore, With such small number statistics, the varying kinematics across multiple IFS samples could be due to varying properties amongst the galaxies selected and instruments used, making it difficult to compare results. 

The global properties of the sample, the most luminous population, studied at $z\sim1.3$ with IFS, are in surprising agreement with a model of unstable disks. 
We agree with \cite{2009ApJ...697.2057L} that statistically all high-redshift galaxies cannot be mergers. Models show that on average the enhancement of SFR in a merger is only a factor of a few and lasts $200-400$ Myr with peak enhancement lasting for less than 100 Myr \citep{2010ApJ...720L.149T,2008A&A...492...31D}. It has been shown locally that all ultra-luminous infrared galaxies (ULIRGS) are mergers \citep{1997A&AS..124..533D} and it is therefore possible a high-redshift subset of galaxies are also exclusively mergers.  However, galaxies on average formed their stars more rapidly in the past implying the need for an evolving definition of what makes more luminous galaxies `extreme'. We therefore stress the importance of the bolometric luminosity in knowing if the sample presented here fall on or above the main sequence of star formation at this redshift to aid in making generalizations about galaxy populations at $z>1$.

Discerning the dust content will aid in a better understanding of their star-formation histories.  
Such moderate to high mass, gas rich, intensely luminous galaxies are reminiscent of local ULIRGs and $z\sim2-3$ SMGs. However, critical to making this link is to determine their far-infrared and sub-mm SEDs, and enabling the measurement of their bolometric luminosities and dust content. Such quantities will inform on whether these galaxies are dust-enshrouded starbursts or Òdust-freeÓ late time proto-galaxies. We will present $Herschel$ data in a future paper that will address the dust content in this sample. 

We expect to see signatures of outflows in these systems as galactic scale winds are expected in galaxies forming stars at $>5$\sfrunits and are seen in 75\% of ULIRGS \citep {2005ARA&A..43..769V}.
Previous studies have detected galactic winds with IFS (high-z: \citealt{Swinbank:2009tw,Alexander:2010bd, Shapiro:2009sj}; local: \citealt{2010ApJ...711..818S}).
The high dispersions to the edges of 02S, 52S, and 02R are real and not an inflation from lower signal-to-noise data. This increase in dispersion towards the edges most likely indicates the presence of galactic winds which add an underlying broad component of \halpha \citep{2005ARA&A..43..769V}. The influence of winds on this sample will be further discussed in a subsequent paper.

Measuring the properties of the individual clumps including size, metallicity, SFR, and velocity dispersion to compare to local samples will be important to testing models of clumpy disk formation. Only on order $\sim10$ clumps have been resolved with IFS studies using lensing and a further $\sim10$ resolved or marginally resolved within the SINS survey \citep{2010arXiv1011.5360G}. The properties of the 10 clumps identified in these systems will be presented in our next paper. \\

In this paper we:\\

\begin{squishitemize}

\item Present kinematic data of 13 star-forming galaxies at $z\sim1.3$, selected from the WiggleZ Dark Energy Survey, of better or comparable resolution than previous studies.

\item Extend the trend of \cite{Green:2010fk} to higher star formation rates and higher dispersions between $1<z<2$ (Figure~\ref{fig.andy}). 

\item Find the most luminous star-forming galaxies at this redshift show a variety of star-forming morphologies with a high incidence of star formation in disks. 

\item Identify clumps of 1--2 kpc in size at $z\sim1.5$ that are resolved kinematically with LGSAO IFS without aid of lensing. 

\item Conclude the clumps in multiple emission galaxies follow the overall kinematics of the system, and find a high incidence of velocity gradients consistent with rotation lending support of the clumpy disk scenario of EBE08 over the major merger scenario.

\end{squishitemize}

\section*{Acknowledgments}
We wish to thank M. Colless for three nights of Directors time on IRIS2. We would also like to thank Jim Lyke and Shelley Wright for their help recreating rectification matrices and Andy Green for helpful discussions on data reduction and kinematics.  We thank the referee for very valuable comments.

We wish to acknowledge Þnancial support from The Australian Research Council (grants DP0772084 and LX0881951 directly for the WiggleZ project, and grant LE0668442 for programming support), Swinburne University of Technology, The University of Queensland, and the Anglo-Australian Observatory. The WiggleZ survey would not be possible without the dedicated work of the staff of the Australian Astronomical Observatory in the development and support of the AAOmega spectrograph, and the running of the AAT.

GALEX (the Galaxy Evolution Explorer) is a NASA Small Explorer, launched in April 2003. We gratefully acknowledge NASAÕs support for construction, operation and science analysis for the GALEX mission, developed in cooperation with the Centre National dÕEtudes Spatiales of France and the Korean Ministry of Science and Technology.

Some of the data presented herein were obtained at the W.M. Keck Observatory, which is operated as a scientific partnership among the California Institute of Technology, the University of California and the National Aeronautics and Space Administration. The Observatory was made possible by the generous financial support of the W.M. Keck Foundation. The authors wish to recognize and acknowledge the very significant cultural role and reverence that the summit of Mauna Kea has always had within the indigenous Hawaiian community.  We are most fortunate to have the opportunity to conduct observations from this mountain.

\bibliographystyle{apj}


\clearpage
\appendix
\label{app}
\section{}
\subsection{Multiple Emission Galaxies}
\subsubsection{WK0912\_13R}
\begin{figure}
\includegraphics[scale=1]{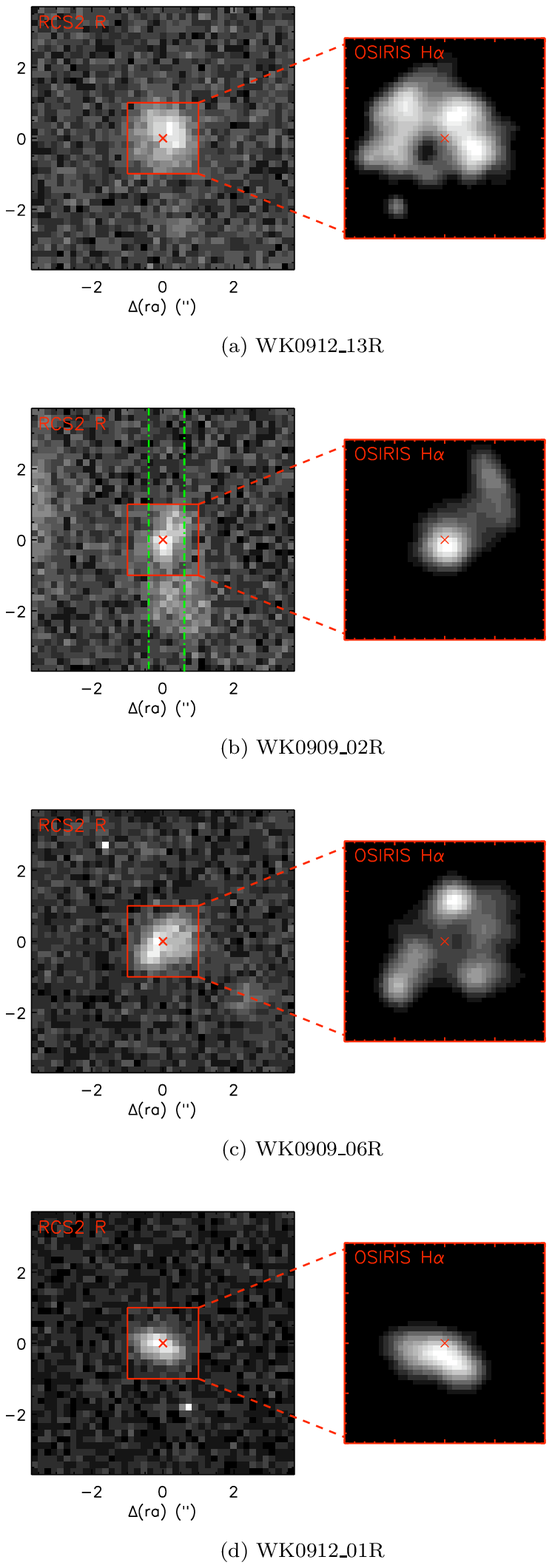}         
\caption{RCS2 r-band and OSIRIS \halpha images for {\it 02R}, {\it 01R}, {\it 06R}, and {\it 13R}, the only objects with resolved broadband imaging. The green dot-dash lines on the RCS2 image of {\it 02R} mark the position of the LRIS long-slit. LRIS spectroscopy confirms the low surface brightness region below the central emission is at the same redshift as the central emission detected by OSIRIS. {\label{fig.RCS2img}}
}
\end{figure}

{\it WK0912\_13R} has an extended irregular flux profile with faint surface brightness at the centre encircled by a brighter ring of emission.  This galaxy shows multiple bright regions of \halpha spatial emission, however due to low signal the individual regions are not resolved. The velocity map shows an ordered velocity gradient with the``hole" in emission spatially placed at the central region of the velocity gradient and is well fit by a disk model. However, this detection has the lowest signal-to-noise of the sample and only a portion of the rotation curve is detected. Optical emission is shown in Figure~\ref{fig.RCS2img}.

\begin{figure}
\includegraphics[scale=1]{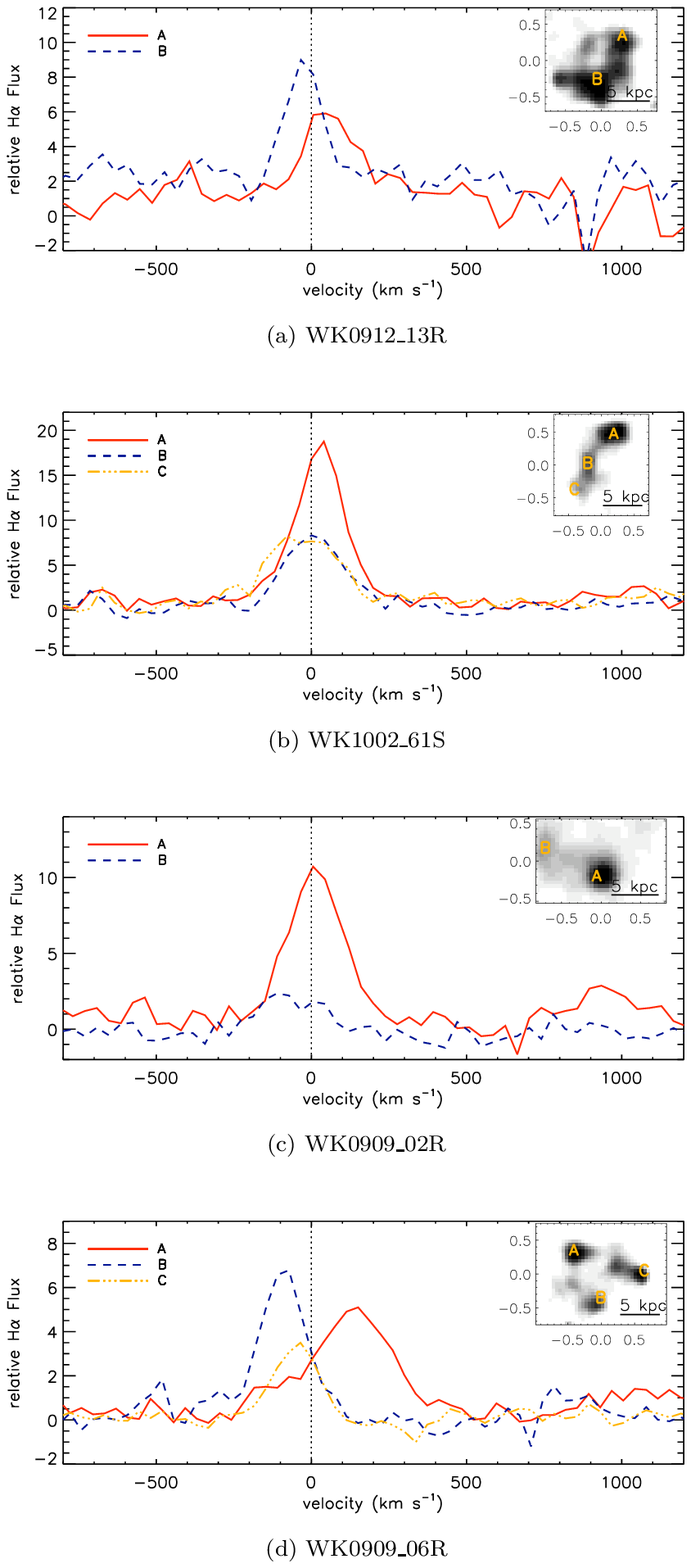}
\caption{ Integrated spectra from individual clumps for all multiple emission systems. Clumps and their corresponding spectra are labeled in \halpha images found in the upper right corner. The spectra are the integrated spectra of emission within 2~r$_{\mathrm{eff}}$ of star-forming regions that are resolved or marginally resolved.
\label{fig.clumps}}
\end{figure}

\subsubsection{WK1002\_61S}
The morphology of {\it61S} reflects that of ``chain galaxies" observed in the Hubble Ultra Deep Field (UDF) with $2-3$ resolved clumps marked by plus signs in Figure~\ref{fig.kinematics}. However, unlike the clumpy galaxies {\it06R} and {\it13R} in this sample, {\it61S} has one of the lowest velocity dispersions of this sample of $\sigma_{\mathrm{mean}}$=85 km s$^{-1}$ and shows little or no velocity separation between clumps as shown in Figure~\ref{fig.clumps}. There is a messy velocity gradient across the length of the ``chain" with peaks of velocity dispersion spatially aligned with the peaks of the eastern most clumps.

\subsubsection{WK0909\_02R}
{\it WK0909\_02R} has the most irregular optical morphology of the objects with available broadband imaging. The RCS2 r-band image (0.185$''$/pixel) shows three distinct regions of emission spread over 4 arcseconds, including the central brightest region denoted by an $\times$ in Figure~\ref{fig.RCS2img}, a lower surface brightness feature to the NW, and a larger low surface brightness feature to the south of the brightest region extending $\sim$21 kpc away from the central peak of broadband emission. 
The OSIRIS observations of {\it WK0909\_02R} has a single peak in \halpha flux with a faint tail of emission to the NW coincident with the peak of r-band emission and faint NW feature respectively. \NII$\lambda\lambda$ 6584~is detected with peak flux coincident with peak \halpha flux.  Long-slit spectroscopy with LRIS reveals the southern low surface brightness component is at the same redshift as the brighter central emission. 
\subsubsection{WK0909\_06R}
{\it WK0909\_06R} has three resolved knots of \halpha emission. In velocity space these peaks correspond to a velocity range of approximately 320 km s$^{-1}$ shown in Figure~\ref{fig.clumps}. This large change in velocity over $\sim$10 kpc leads to broad velocity width of the \halpha emission line between flux peaks where a signature of two Gaussians is seen. 
Peak \NII~emission corresponds with the brightest peak of \halpha and is not detected elsewhere in the galaxy.  Faint \SII~emission is detected in the integrated spectrum but does not have high enough signal-to-noise to be mapped spatially. 
The RCS2 image in Figure~\ref{fig.RCS2img} reveals a possible optical companion 1.8$''$ to the south-west (15 kpc). Future long-slit spectroscopy will determine if the galaxies are under-going a merger.

\subsection{Extended Emission Galaxies}
\subsubsection{WK0905\_22S}
The flux profile of {\it WK0905\_22S} is also best fit by two components with the  the primary peak of velocity dispersion occurring at the boundary of the two systems indicative of a line-of-sight merger, however the velocity map shows evidence of rotation and is well fit by a disk.

\subsubsection{WK0912\_01R}
{\it WK0912\_01R} has one extended peak of \halpha flux with a fainter surrounding halo. The velocity map shows a strong velocity gradient of \vshear$\sim200$ km s$^{-1}$ and a near linear rotation curve. It is well-fit by a disk with optical morphology consistent with being an edge-on system as seen by the ellipticity of the rest-frame UV image in Figure~\ref{fig.RCS2img}.
\NII~ was not detected in the OSIRIS observation of this galaxy, which is partly due to its short integration time, an upper limit is given in Table~\ref{flux.table}.

\subsubsection{WK1002\_41S}
The \halpha morphology of {\it41S} reveals two peaks of emission connected by a dip in the \halpha surface brightness at the central region of the detection. The rotation profile of {\it 41S} is linear ($v_{\mathrm{shear}}$=114 km s$^{-1}$) with a featureless velocity dispersion; as a result there exist a strong degeneracy between $r_\mathrm{p}$ and $V_\mathrm{p}$ when fitting a disk profile. This indicates that {\it41S} is well-fit by a disk but as we only detect the inner part of rotation, the kinematic extent cannot be constrained. It follows that the dispersion map is flat and featureless as the turnover has not been reached. 

\subsubsection{WK0809\_02S}
The 2-D flux profile of {\it WK0809\_02S} is best fit by two emission components. The peak in velocity dispersion corresponds with the brightest peak in flux, however no clear velocity gradient is seen and the object has the lowest velocity shear in the sample (\vshear=36 km s$^{-1}$). With $J$ band photometry ($K$ not available) the stellar mass is estimated at $\log(\mathrm{M_{*}})$ = $11.1\pm0.3$~\Msun, one of the most massive of the sample. A tentative IRAM detection of the CO(2-1) emission line exists at 500 km s$^{-1}$ from the $z_{\mathrm{sys}}$ of \halphans.  The CO(2-1) line is formally 2.5$\sigma$ per channel, but spans $>$3 channels, so is $\sim$5$\sigma$ integrated over the line, possibly indicative of an outflow/inflow signature. This result is outside of the scope of this analysis and will be addressed in a subsequent paper.

\subsection{Single Emission Galaxies}

\subsubsection{WK1002\_18S}
{\it WK1002\_18S} is a compact resolved single emission galaxy with $n_{\mathrm{H}\alpha}$=0.8 and shows a near-linear velocity gradient. The stellar mass of {\it18S} is $8\times10^{9}$\Msun and has a velocity dispersion \smean$=102$ km s$^{-1}$.

\subsubsection{WK1002\_14S}
{\it WK1002\_14S} has a single compact region of resolved \halpha emission showing a linear velocity shear, with some variations in the velocity field at the edges of the galaxy detection. As with {\it41S} there exists a degeneracy between $r_\mathrm{p}$ and $V_\mathrm{p}$ due to the linear nature of the velocity profile. The flat dispersion map can be explained by the lack of turn-over in the velocity profile. This galaxy is one of the highest dispersion galaxies with \smean$\sim$130 km s$^{-1}$ and stellar mass of M$_{*}$=$3.7\times10^{10}$ \Msun.  \NII $\lambda\lambda6583.5$ is detected coincident with \halpha emission. 

\subsubsection{WK1002\_46S}
{\it WK1002\_46S} is the most compact ( $r_{1/2}$(\halphans)$\sim$1.8 kpc ) resolved galaxy. The stellar mass of {\it46S} is $7\times10^{9}$\Msun and the galaxy has a relatively low \smean$=85$ km s$^{-1}$.

\subsubsection{WK1002\_52S}
{\it 52S} is only marginally resolved and an outlier in many respects including: highest velocity dispersion (\smean=125 km s$^{-1}$), highest stellar mass (M$_{*}$=$4.6\times10^{11}$ \Msun), most compact ($r_{1/2}\sim$1.8 kpc). Given the high stellar mass it is more likely that 52S is a merger remnant rather than a newly forming bulge.

\subsubsection{WK0912\_16S}
{\it WK0912\_16S} is observed to have a single resolved peak in \halpha flux. \NII~is detected with peak flux coincident with the peak \halpha flux. {\it16S} has one of the lowest velocity dispersions of the sample at \smean$\sim90$ km s$^{-1}$, and does not show a coherent velocity gradient. It also has the lowest stellar mass of the sample of $6.3\times10^{9}$ \Msun~ and the flattest flux profile with $n_{\mathrm{H}\alpha}$=0.5$^{+0.3}_{-0.1}$.

\label{lastpage}

\end{document}